\documentclass[11pt]{article}

\usepackage[T1]{fontenc}
\usepackage[cp1251]{inputenc}
\usepackage{textcomp}
\usepackage[centertags]{amsmath}
\usepackage{amsfonts}
\usepackage{amssymb}
\usepackage{enumerate}
\usepackage{bbold}
\usepackage[numbers,sort&compress]{natbib}
\usepackage[hypertex]{hyperref}

\usepackage{paperinitial}

\paperinitialization{15mm}{15mm}{20mm}{10mm}{2pt}{10pt}

\DeclareMathOperator{\pr}{pr}
\DeclareMathOperator{\re}{Re}
\DeclareMathOperator{\im}{Im}

\newcommand{\lan}{\langle}
\newcommand{\ran}{\rangle}

\newcommand{\e}{\varepsilon}
\newcommand{\vf}{\varphi}
\newcommand{\vk}{\varkappa}
\newcommand{\s}{\sigma}

\newcommand{\al}{\alpha}
\newcommand{\be}{\beta}
\newcommand{\ga}{\gamma}
\newcommand{\Ga}{\Gamma}
\newcommand{\de}{\delta}
\newcommand{\De}{\Delta}
\newcommand{\ka}{\varkappa}
\newcommand{\la}{\lambda}
\newcommand{\La}{\Lambda}
\newcommand{\ups}{\upsilon}

\newcommand{\spk}{\mathbf{k}}

\begin{document}
\allowdisplaybreaks[4]
\title{{\Large \textbf{Quantum gravitational anomaly as a dark matter}}}

\date{}

\author{P.~O. Kazinski\thanks{E-mail: \texttt{kpo@phys.tsu.ru}}\\[0.5em]
{\normalsize Physics Faculty, Tomsk State University, Tomsk 634050, Russia}}

\maketitle

\begin{abstract}

The general properties of a perfect relativistic fluid resulting from the quantum gravitational anomaly are investigated. It is found that, in the limit of a weak gravitational field, this fluid possesses a polytropic equation of state characterized by two universal constants: the polytropic constant and the natural polytropic index. Based on the astrophysical data, the estimates for the polytropic constant are given. It is shown that this fluid can describe a considerable part of the cold dark matter. The quantum theory of such a fluid is constructed in the framework of the background field method. The Ward identities associated with the entropy and vorticity conservation laws are derived. The leading gradient corrections to the pressure of the perfect fluid are found and the restrictions on their form are obtained. These restrictions guarantee, in particular, the absence of ghosts in the model. The second order nonlinear corrections to the equations of motion of a perfect relativistic fluid are analyzed and the explicit expressions for the transverse and longitudinal perturbations induced by a sufficiently strong sound wave are obtained. A dynamical solution to the problem of time in quantum gravity is proposed.

\end{abstract}

\section{Introduction}

There are many approaches to explain the general features of the phenomenon of a cold dark matter. The common point of view is that the dark matter is a gas of weakly interacting massive particles (see, e.g., \cite{GorbRub1,Schneidr}); another one is that the dark matter is a manifestation of deviations of the gravitational laws from that dictated by General Relativity (see, e.g., \cite{Odints}); the third approach postulates a modification of dynamics at small accelerations or introduces the additional tensor fields into the model (see \cite{Milgr,Sanders,FamaMcGau}); and so on. In the series of papers \cite{olep,KazShip,gmse,prop,KalKaz1,KalKaz2}, it was proved that the quantum gravitational anomaly exists\footnote{This anomaly is not of the type found in \cite{AlvaWitt}.}. Its existence leads in a natural way to the contribution to the energy-momentum tensor with the properties resembling a cold dark matter \cite{gmse}. Such a dark matter is described by the equations of relativistic hydrodynamics with a certain equation of state. In this paper, we shall study the general properties of this fluid, with the assumption that it is responsible for the most part of a dark matter. We shall also construct its quantum theory using the background field method \cite{DeWGAQFT,BuchOdinShap}. The latter solves in a dynamical manner the issue of dependence of the quantum field theory observables on a choice of the vacuum state for an arbitrary gravitational background. In other words, this solves the so-called problem of time in quantum gravity \cite{DeWQG1,DeWQG2,Isham,IshKuch,KuchTor,KuchBrow,ConnRov,GriMaMos,MamMostStar}.

As known, besides the perturbative nonrenormalizability, quantum gravity possesses another one important problem -- its precise formulation depends on a choice of the vacuum state for quantum fields. Different vacuum states result in different, and in many cases unitary inequivalent, quantum theories. For example, it is notorious the dependence of the effective action on a vacuum state in black hole physics \cite{GriMaMos,DeWQFTcspt,DeWGAQFT,BirDav,FrolNov}. Usually, the three types of vacuum states are distinguished for quantum fields evolving on the black hole backgrounds: the Boulware vacuum \cite{Boulware}, the Hartle-Hawking vacuum \cite{HartHawk,Israel}, and the Unruh vacuum \cite{Unruh}. However, it is clear that one can construct infinitely many vacua leading to infinitely many inequivalent quantum field theories. The same effect, but formulated in other terms, arises as the time problem in quantum gravity \cite{Isham,IshKuch,KuchTor,KuchBrow,ConnRov}.

A unitary inequivalence of quantum field theories having the same classical limit reveals itself in the infrared and ultraviolet divergencies in the effective action. Of course, some divergencies appearing in the perturbative calculations may cancel upon summing an infinite number of diagrams, but the complete cancelation of divergencies in the model without any large symmetry, for instance, in the standard model, is improbable. Therefore, one may regard the perturbative divergencies as a measure of a unitary inequivalence. The choice of the regularization scheme and concrete values of the constants at the singular structures (by introducing the counterterms to the initial classical action) defines a unique representation of the algebra of observables. For example, the standard ultraviolet divergencies in quantum field theory correspond to a unitary inequivalence of the Heisenberg representation and the interaction picture \cite{Haagb}. For the renormalizable theories, a passage from one representation to another is reduced to a renormalization of the coupling constants\footnote{Usually, one considers a class of the unitary inequivalent theories possessing certain symmetries: the Poincar\'{e} invariance, the gauge invariance, etc.} (see, e.g., \cite{DiEfGaNe}).

On including gravity into quantum field theory, the divergencies become dependent on the vector field $\xi^\mu$ (see, e.g., \cite{DeWQFTcspt,olep,gmse,KalKaz1}) defining the physical regularization \cite{CollinsPhys,ColPerSud,Collins} by the energy cutoff, determining the Hamiltonian of quantum fields and the vacuum state. If one appeals to the Gelfand-Naimark-Segal (GNS) construction then this vector field specifies the vacuum state and, consequently, the representation of the algebra of observables (see, e.g., \cite{Emch,ConnRov}). Of course, one can choose a certain unique representation fixing the vector field $\xi^\mu$ by hand as it is done in describing the quantum effects near black holes. However, the presence of the non-dynamical field $\xi^\mu$ in the effective action leads inevitably to a quantum gravitational anomaly, i.e., to a violation of the general covariance (the background independence). One of the possible solutions to this problem is to define $\xi^\mu$ dynamically. Then the general covariance is restored, but at the cost of the additional degrees of freedom. It turns out \cite{gmse} that the dynamics of the field $\xi^\mu$ are unambiguously determined from the requirement of the fulfillment of the Ward identities for the effective action and are described by the hydrodynamic equations of motion. We shall take into account these observations in the condition 3 of Sect. \ref{QGA} concerning the structure of the classical action for the field $\xi^\mu$.

From the renormalization theory viewpoint (see, e.g., \cite{BogolShir,Collins}), one could suppose that the divergencies mentioned above ought to be simply canceled out by the appropriate counterterms, and, thereby, the general covariance could be restored in the effective action. The principal result of the paper \cite{KalKaz2} consists in the proof that the effective action of quantum gravity contains the finite contributions that are not expressed in a covariant form in terms of the metric alone and cannot be canceled by the counterterms. Even the explicit expressions for these contributions were found in \cite{KalKaz2} for a stationary slowly varying in space gravitational background. These contributions are not analytic in the gravitational constant and momenta. The fact that such terms should be present in the effective action was frequently mentioned in the literature (see, e.g., \cite{GriMaMos,ZelnProc,GavrGit}). The non-covariant terms analytic in the gravitational constant and momenta also appears in calculating the quantum effects on the black hole backgrounds \cite{Page,BrOtPa,FrZel,FrolNov,AndHisSam,Howard}. However, in the latter case, it was not clear whether these contributions must be canceled by the counterterms or not. The presence of the non-covariant terms in the effective action means that the general covariance is violated in quantum gravity. This anomaly is characterized by the vector field $\xi^\mu$ referred above, that is, the effective action can be made formally covariant, if one introduces this vector field \cite{FrZel,FrolNov,AndHisSam}. Therefore, following the general prescriptions of renormalization theory \cite{Collins} and of the effective field theory approach \cite{Weinb,DonGolHol}, we have to include $\xi^\mu$ into the initial classical action of a theory and make this field dynamical. This, in particular, implies that a complete cancelation to zero of all the divergencies depending on $\xi^\mu$ is unnatural. The various examples of quantum field theory models teach us that the structures having a small number of derivatives and appearing as divergencies constitute the classical action.

The fact that the presence of the vector field $\xi^\mu$ in the effective action is revealed as the presence of some perfect fluid resembling a cold dark matter in the weak gravitational field limit was mentioned in \cite{gmse,prop,KalKaz1,KalKaz2}. In this paper, we develop this proposition. In Sect. \ref{QGA}, we briefly formulate the main properties of the quantum gravitational anomaly. In Sect. \ref{FTFormRH}, the Fock-Taub formalism for relativistic hydrodynamics resulting from the gravitational anomaly is expounded. Then, in Sect. \ref{QuantHyd}, we discuss the quantization of relativistic hydrodynamics employing the background field method. Actually, we study the dynamics of small fluctuations over the classical background. In Sects. \ref{ClassPot} and \ref{PolEoS}, we establish the general properties of the fluid pressure coming from the gravitational anomaly. In particular, we show that this fluid possesses the polytropic equation of state with a universal polytropic constant and a universal natural polytropic index. In the next Sects. \ref{GradCorr}, \ref{LinEqMot}, and \ref{StabCrit}, the influence of the leading gradient corrections to the fluid pressure on the dynamics of small perturbations (the phonons) is investigated and the restrictions on the form of these corrections are found. Also, we derive in these sections the explicit form of the Ward identities associated with the entropy and vorticity potential conservation laws. Sect. \ref{NonlinCorr} is devoted to the leading nonlinear corrections to the dynamics of the longitudinal and transverse phonons. In particular, we find the explicit expression for the transverse perturbations induced by a longitudinal sound wave. Then, in Sect. \ref{ConFormClPot}, we consider a concrete form of the fluid pressure that follows from the renormalization theory. In Sect. \ref{QuanEff}, we discuss some quantum effects induced by the phonons and show that the standard in-out perturbation theory has a very narrow range of applicability in the weak field limit \cite{ENRW}.

We shall use the conventions adopted in \cite{DeWGAQFT}
\begin{equation}
    R^\al_{\ \be\mu\nu}=\partial_{[\mu}\Ga^\al_{\nu]\beta}+\Ga^\al_{[\mu\ga}\Ga^\ga_{\nu]\beta},\qquad R_{\mu\nu}=R^\al_{\ \mu\al\nu},\qquad R=R^\mu_\mu,
\end{equation}
for the curvatures. The square and round brackets at a pair of indices denote antisymmetrization and symmetrization without $1/2$, respectively. The Greek indices are raised and lowered by the metric $g_{\mu\nu}$ which has the signature $-2$. The system of units is chosen such that $c=\hbar=1$.

\section{Quantum gravitational anomaly}\label{QGA}

In this small section, we summarize the results of the papers \cite{olep,KazShip,gmse,prop,KalKaz1,KalKaz2} necessary for us. It was shown in this series of papers that, in the background field gauge \cite{DeWGAQFT,BuchOdinShap}, the renormalized in-in effective action of the fields $\Phi$ and $g_{\mu\nu}$ has the form (see, for details, \cite{gmse})
\begin{equation}
    \Ga[g^+_{\mu\nu},\xi^+_\mu,\Phi^+;g^-_{\mu\nu},\xi^-_\mu,\Phi^-],
\end{equation}
where $\Phi$ denotes all the matter fields including the Faddeev-Popov ghosts both for the matter and metric fields. The dependence on the field $\xi^\mu$ characterizes the dependence of the effective action on a choice of the vacuum state for quantum fields (for instance, using the GNS construction) and, in particular, the regularization procedure.

The requirement of invariance of the effective action with respect to general coordinate transformations implies the Ward identity
\begin{equation}\label{Ward_id}
    \nabla^\nu_+ T_{\mu\nu}\approx\mathcal{L}_{\xi^+}\Ga_\mu+\nabla^\nu_+\xi^+_\nu\Ga_\mu,\qquad \sqrt{|g^+|}T^{\mu\nu}:=-2\frac{\de\Ga}{\de g^+_{\mu\nu}},\quad\sqrt{|g^+|}\Ga_\mu:=-\frac{\de\Ga}{\de\xi^{+\mu}}.
\end{equation}
The plus sign at the connection means that it is constructed using the metric $g^+_{\mu\nu}$. The approximate equality denotes  the fulfilment of the Ward identity on the solutions to the quantum equations of motion of the fields $\Phi$. Henceforth, for brevity, we omit the plus sign at the ``$+$'' fields. As we see, the Ward identity for the matter fields, i.e., a vanishment of any matrix element of the energy-momentum tensor divergence operator, is violated. For it is satisfied, one needs to demand
\begin{equation}\label{Ward_id_ii}
    \mathcal{L}_{\xi}\Ga_\mu+\nabla^\nu\xi_\nu\Ga_\mu=0.
\end{equation}
These are the equations of a hydrodynamical type for the field $\xi_\mu$. They can be rewritten as
\begin{equation}\label{eqmot xi}
    \nabla_\mu(\xi^\mu w)=0,\qquad \xi^\mu\nabla_{[\mu}(\Ga_{\nu]}/w)=\mathcal{L}_\xi(\Ga_{\nu}/w)=0,
\end{equation}
where $w:=\xi^\rho\Ga_\rho$. The same equations hold for the field $\xi^-_\mu$ too. In other words, the regularized operator equality
\begin{equation}\label{Ward_id_iii}
    \hat{\nabla}^\nu \hat{T}_{\mu\nu}[\hat{g}_{\mu\nu},\hat{\xi}^\mu,\hat{\Phi}]=0,
\end{equation}
where all the quantum fields are expressed in terms of the creation-annihilation operators, should be the Heisenberg equation of motion of the quantum field $\hat{\xi}^\mu$.

Despite the fact that, in using the standard procedure to calculate the averages of quantum operators, the dependence on the field $\xi^\mu$ appears only in the quantum corrections to the effective action, the general scheme of renormalization theory (see, e.g., \cite{Collins}) says that the field $\xi^\mu$ must be included into the initial classical action as well. The equations of motion for this field should be equivalent to the condition of the covariant divergenceless of the energy-momentum tensor. In this case, a consistent quantization of the fields $g_{\mu\nu}$, $\Phi$, and $\xi^\mu$ results in the quantum dynamics preserving the Ward identity \eqref{Ward_id_ii}, \eqref{Ward_id_iii}. In constructing the classical Hamiltonian of the system and in introducing the physical regularization (the energy cutoff), the field $\xi^\mu$ ought to be used. This field is an average of the operator $\hat{\xi}^\mu$. As expected, we see that the violation of the Ward identity leads to new degrees of freedom in the theory -- the dynamical field $\xi^\mu$.

The natural normalization conditions are imposed on the effective action involving the field $\xi^\mu$:
\begin{enumerate}[1)]
  \item The Lorentz-invariance. For the flat space-time, where $g_{\mu\nu}=\eta_{\mu\nu}$, the dependence of the effective action on the field $\xi^\mu$ should disappear;
  \item The compliance with the Einstein equations. In the flat space-time limit, the vacuum expectation value of the energy-momentum tensor operator of the matter fields and the field $\xi^\mu$ must be zero;
  \item Minimality. The initial classical action of the field $\xi^\mu$ contains only those structures that arise as divergencies in calculating the quantum corrections using the physical regularization (see, e.g., \cite{CollinsPhys,ColPerSud,Collins}) by the energy cutoff. It is these terms that specify the energy-time anomaly \cite{prop,KalKaz1} and do not include the monomials invariant with respect to the global dilatations of $\xi^\mu$.
\end{enumerate}
The last requirement implies, in particular, that, in the flat space-times limit, the initial classical action of the field $\xi^\mu$ should contain only renormalizable vertices provided the theory of the fields $\Phi$ is renormalizable in a flat space-time.

These requirements restrict the form of the classical action of the field $\xi^\mu$, but, unfortunately, do not fix it unambiguously. When the energy cutoff regularization is employed, the regularization parameter $\La$ always enters the effective action in the form of the combination $\xi^2/\La^2$. Therefore, if the theory does not contain the coupling constants with a negative mass dimension as, for example, the standard model without gravity, then it follows from the third requirement above that the coupling constants at the vertices in the action of the field $\xi^\mu$ possess nonnegative mass dimensions provided the field $\xi^\mu$ has been stretched as $\xi^\mu\rightarrow\La\xi^\mu$. The inclusion of the gravitational corrections makes this statement improper (see, e.g., \cite{Ventwoloop}). Nevertheless, if $\La\ll M_{Pl}$, $p\ll\La$, where $M_{Pl}$ is the Planck mass and $p$ is a characteristic momentum flowing in the Feynman diagrams, then these gravitational corrections are small and can be taken into account systematically using the effective field theory techniques \cite{Weinb,DonGolHol}. In the papers \cite{DeWQFTcspt,gmse,KalKaz1}, the one-loop divergencies were found. Their structure agrees with the property mentioned above. Note that the dilatation $\xi^\mu\rightarrow\La\xi^\mu$ allows one to get rid of the regularization parameter $\La$ in the theory. In that case, it is replaced by the field $\xi^\mu$ with the mass dimension $-1$.

In order to impose the additional restrictions on the form of the effective action and, consequently, in the leading order of the perturbation theory, on the form of the classical action of the field $\xi^\mu$, one can demand the fulfilment of the conformal (Weyl) invariance of the effective action on the cutoff scale $\La$. In virtue of the fact that the conformal and the energy-time anomalies are related \cite{KalKaz1}, this condition fixes uniquely the coefficients at the logarithms (see \cite{KalKaz1} and \eqref{eff_pot} below) in the effective action of the field $\xi^\mu$: these coefficients should coincide with or be proportional to the renormalization group scaling functions $\be$ and $\ga$. However, these renormalization group functions are calculated on the scale $\La$, and all the particles with masses $m\ll\La$ make contributions to the values of these functions. As long as we do not know whether the standard model is a final one or there exist more massive particles that are not included into it, these scaling functions are unknown. So, this restriction on the coefficients of the action for the field $\xi^\mu$ is not constructive in the case when the complete Hamiltonian of the theory is hidden. Even if the theory were completely given, it could become non-perturbative on the scale $\La$, and the standard methods to determine the scaling function would be inapplicable. Therefore, in this paper, imposing additional constraints on the coefficients at the terms in the action of the field $\xi^\mu$, we shall proceed from the empirical reasonings adjusting these coefficients in such a way that they provide the required asymptotics of the energy-momentum tensor in the weak field limit.

We shall discuss in details the form of the classical potential of the field $\xi^\mu$ in Sects. \ref{ClassPot} and \ref{ConFormClPot}. At the present stage, it is sufficient for our purposes that this potential obeys the conditions 1 and 2. The explicit form of the gradient terms and their influence on the dynamics of the field $\xi^\mu$ will be considered in Sect. \ref{GradCorr}.

\section{Fock-Taub formalism for relativistic hydrodynamics}\label{FTFormRH}

As we have seen in the previous section, we need to construct a model of quantum relativistic hydrodynamics in order to obtain a self-consistent quantum field theory involving the quantum gravitational field. At the present moment there are many examples of quantization of relativistic hydrodynamics in the literature (see, e.g., \cite{ENRW,DHNS,Torri,BallBell,GripSuth,Ballest}). However, these approaches are based on the classical action different from that dictated by the quantum gravitational anomaly. It is well-known that, upon quantization, different action functionals leading to the same equations of motion may result in different quantum theories even for the models with finite degrees of freedom (see, e.g., \cite{KLSh}). The Fock-Taub formalism \cite{FockB,Taub} is the most appropriate (and seems to be unique) to treat relativistic hydrodynamics stemming from the quantum gravitational anomaly: the fluid flow is characterized by a single vector field $\xi^\mu$, and the variation of the action leads to the equations of motion equivalent to the condition of the covariant divergenceless of the energy-momentum tensor \eqref{eqmot xi}. Other Lagrangian descriptions of relativistic hydrodynamics can be found, for example, in \cite{Schutz,Ray,KiSmGo,Brown1,Brown2,HajKij,rrmm}.

Let $x^\mu(\vk)$, $\mu=\overline{0,3}$ and $\vk=\{\tau,\s^i\}$, $i=\overline{1,3}$, be a orientation preserving map of the four-dimensional manifold $N$ to a four-dimensional space-time $M$ equipped with the metric $g_{\mu\nu}$. Let $\rho^a(\vk)$, $a=\overline{0,3}$, be a vector field on $N$ such that $\rho^ah_{ab}\rho^b>0$, where $h_{ab}:=\partial_a x^\mu\partial_b x^\nu g_{\mu\nu}$ is the induced metric on $N$. Then the action functional \cite{FockB,Taub} (cf. \cite{ENRW,DHNS,Torri,BallBell,GripSuth,Ballest}),
\begin{equation}\label{action_FT}
    S[x(\vk)]=\int_N d\vk\sqrt{|h|}p[\rho^a,h_{ab}],
\end{equation}
describes perfect isentropic relativistic hydrodynamics in the Fock-Taub representation. Here $p$ is some scalar constructed in terms of $\rho^a$ and $h_{ab}$ and their derivatives. As a matter of fact, it also depends on the fields $\Phi(x(\vk))$, but this will be irrelevant for our further considerations provided the fields $\Phi(x(\vk))$ obey their equations of motion. Note that $\rho^a$ is not a dynamical field. It is not varied in taking a variation with respect to $x(\vk)$.

Since the action depends on $x(\vk)$ only through the metric $h_{ab}$ and the fields $\Phi$, we have
\begin{equation}\label{action_variat}
    \frac{\de S}{\de x^\mu(\vk)}=\int d\vk'\det[e_a^\mu(\vk')]\frac{\de S}{\de g_{\la\rho}(x(\vk'))}e^a_\la(\vk') e^b_\rho(\vk')\frac{\de h_{ab}(\vk')}{\de x^\mu(\vk)}
    =\det e_a^\mu\Big[ -2\partial_\la\Big(\frac{\de S}{\de g_{\la\rho}}g_{\mu\rho}\Big) +\frac{\de S}{\de g_{\la\rho}}\partial_\mu g_{\la\rho}\Big],
\end{equation}
on the solutions to the equations of motion of the fields $\Phi$. The notation $e_a^\mu:=\partial_ax^\mu$ has been introduced, and we have used the relation
\begin{equation}
    \partial_b(e^b_\rho\det e_a^\mu )=0.
\end{equation}
The vectors $e_a^\mu$ constitute a holonomic basis, but not a tetrad. The expression in the square brackets in \eqref{action_variat} is expressed via the divergence of the energy-momentum tensor defined in the standard manner \eqref{Ward_id}. Hence, we arrive at
\begin{equation}\label{hydr_eqm00}
    \frac{\de S}{\de x^\mu(\vk)}=\det e_a^\mu\sqrt{|g|}\nabla_\la T^\la_\mu=\sqrt{|h|}\nabla_\la T^\la_\mu=0,
\end{equation}
i.e., the equations of motion following from the action \eqref{action_FT} are equivalent to the condition of covariant divergenceless of the energy-momentum tensor constructed by the use of this action functional.

Putting $\xi^\mu=\rho^ae_a^\mu$, we obtain the action functional for the field $\xi^\mu$ discussed in the previous section. As long as
\begin{equation}
    \nabla_\mu e^\nu_a=e^b_\mu(\partial_b e^\nu_a-\Ga_{ba}^ce_c^\nu)+\Ga^\nu_{\mu\rho}e^\rho_a=0,
\end{equation}
where $\Ga_{ba}^c$ is the Levi-Civita connection constructed in terms of the metric $h_{ab}$, we deduce from \eqref{hydr_eqm00},
\begin{equation}\label{hydr_eqm_0}
    e^\mu_a\frac{\de S}{\de x^\mu(\vk)}=:\sqrt{|h|}E_a=\sqrt{|h|}\nabla_b T^b_a=0.
\end{equation}
Performing an infinitesimal general coordinate transform of the variables $\ka$ in the action \eqref{action_FT}, we prove, as in the previous section, that
\begin{equation}\label{hydr_eqm}
    \nabla^b T_{ab}=\mathcal{L}_{\rho}S_a+\nabla^b\rho_bS_a=0\;\;\Leftrightarrow\;\;\nabla_a(\rho^a w)=0,\qquad \mathcal{L}_\rho(S_a/w)=0,
\end{equation}
where $S_a$ is defined similarly to $\Ga_\mu$ and the enthalpy density $w=\rho^a S_a$. If one defines the reciprocal temperature and entropy density as \cite{gmse}
\begin{equation}\label{temp_entrop}
    \be:=\sqrt{\rho^2}, \qquad \s:=\be w,
\end{equation}
then the first equation in the system \eqref{hydr_eqm} is the entropy conservation law for a perfect fluid. It is obtained by convolving with $\rho^a$ the condition of covariant divergenceless of the energy-momentum tensor.

The second equation in the system \eqref{hydr_eqm} implies the vorticity conservation law. The one-form,
\begin{equation}\label{genT}
    r_a:=S_a/w,\qquad\rho^a r_a=1,
\end{equation}
is conserved along the fluid flow according to the second equation in \eqref{hydr_eqm}. For a particular case, when the Lagrangian density of the action \eqref{action_FT} does not contain higher derivatives of the fields $x^\mu(\vk)$ (see, for instance, \eqref{eff_pot}), the one-form \eqref{genT} is equal to the Tolman temperature one-form $t_a=\rho_a/\rho^2$ \cite{gmse}. The vorticity tensor can be defined as (see, e.g., \cite{Mitskev})
\begin{equation}\label{vorticity}
    A_{ab}:=\pi_a^c\pi_b^d\partial_{[c}r_{d]}=(dr+\mathcal{L}_\rho r\wedge r)_{ab},\qquad\pi_a^b:=\de_a^b-\rho^br_a.
\end{equation}
It is zero if and only if the one-form $r_a$ is integrable. On the solutions to the equations of motion \eqref{hydr_eqm}, the vorticity reduces to the external derivative of the one-from $r_a$ and so the latter can be called the vorticity potential. Owing to the property of the Lie derivative, the second equation in \eqref{hydr_eqm} entails the vorticity tensor conservation.

In the adapted system of coordinates, where the vector field  $\rho^a$ is straighten [$\rho^a=(1,0,0,0)$], the action functional \eqref{action_FT} is invariant under the symmetry transformations (see, e.g., \cite{LandLifshCTF,Zelman,Vladimir,Moller})
\begin{equation}\label{symm_hydr}
\begin{gathered}
    \tau\rightarrow\tau'=\tau+f(\s),\qquad \s\rightarrow\s'=\s'(\s);\\
    \de_\e\rho^a=0,\qquad\de_\e x^\mu(\tau,\s)=\e^a(\s)\partial_a x^\mu(\tau,\s),
\end{gathered}
\end{equation}
for any fixed metric $g_{\mu\nu}(x)$. Notice that these local transformations are independent of time $\tau$ and so are not gauge transformations (see the definition in \cite{HeTe}). Applying this symmetry transform to the action \eqref{action_FT}, we come to the relation
\begin{equation}\label{vort_cons_law}
    \sqrt{|h|}E_a=\partial_\tau f_a[x(\vk)],
\end{equation}
where $f_a[x(\vk)]$ is a certain local expression. Consequently, we have the conservation laws provided the equations of motion are fulfilled. Comparing it with \eqref{hydr_eqm} in the adapted system of coordinates, we see that the zeroth component of the conservation law is responsible for the entropy conservation, while the spatial components are for the $r_a$  conservation. Notice that the zeroth component of the second conservation law in \eqref{hydr_eqm} is trivial in this system of coordinates.

\subsection{Hamiltonian}

In order to construct quantum relativistic hydrodynamics, we shall not use the canonical formalism explicitly. In our case, it is rather awkward since the momenta cannot be expressed through the velocities in an explicit form, although it can be done in principle (see for details \cite{Brown1,Brown2}). Quantizing relativistic hydrodynamics, we shall use the background field method \cite{DeWGAQFT,BuchOdinShap} which states that the knowledge of quantum evolution of small fluctuations over an arbitrary background is sufficient to reconstruct the whole quantum dynamics. Nevertheless, it is important to realize what the canonical Hamiltonian for the action \eqref{action_FT} is. Naively, one may expect that it is equal to the energy of a fluid. However, we shall see that it is not so for the fields $x(\vk)$.

For simplicity, we restrict our considerations to the case when the Lagrangian density of the action \eqref{action_FT} is independent of the higher derivatives (of the second order and higher) of the fields $x^\mu(\vk)$ with respect to $\tau$. Then the density of the canonical Hamiltonian takes the form
\begin{equation}
    H_{can}[x(\vk)]=\dot{x}^\mu\frac{\partial}{\partial \dot{x}^\mu}(\sqrt{|h|}p)-\sqrt{|h|}p,\qquad \dot{x}^\mu:=\partial_\tau x^\mu.
\end{equation}
Introducing the spatial metric \cite{LandLifshCTF,Zelman,Vladimir,Mitskev}
\begin{equation}
    q_{ij}:=h_{ij}-\frac{\dot{x}^\mu g_{\mu\nu}e_i^\nu \dot{x}^\la g_{\la\rho}e_j^\rho}{\dot{x}^\mu g_{\mu\nu}\dot{x}^\nu},\qquad\det{h_{ab}}=\dot{x}^2\det q_{ij},
\end{equation}
we obtain
\begin{equation}
    H_{can}[x(\vk)]=\sqrt{|q|}\sqrt{\dot{x}^2}\dot{x}^\mu\frac{\partial p}{\partial \dot{x}^\mu}.
\end{equation}
The expression $\dot{x}^\mu\partial p/\partial \dot{x}^\mu$ determines a variation of $p$ under the action of an infinitesimal dilatation of the variable $\dot{x}^\mu$. The local dilatation transform $\dot{x}^\mu\rightarrow\la(\vk) \dot{x}^\mu$ acts on the metric $h_{ab}$ as
\begin{equation}
    h_{ab}\rightarrow h'_{ab}=J^c_a h_{cd}J^d_b,\qquad J=\left[
                                                           \begin{array}{cc}
                                                             \la(\vk) & 0 \\
                                                             0 & \mathbb{1}_{3\times 3} \\
                                                           \end{array}
                                                         \right].
\end{equation}
As long as $p$ is a scalar, we have
\begin{equation}
    p[\rho^a,h'_{ab}]=p[J^a_b\rho^b,h_{ab}]=p[\la(\vk)\rho^a,h_{ab}],
\end{equation}
where the last equality is valid in the adapted system of coordinates. Therefore,
\begin{equation}
    S[\la(\vk)\rho]-S[\rho]=\int d\vk \sqrt{|h|}(p[\la(\vk)\rho^a,h_{ab}]-p[\rho^a,h_{ab}])\approx\int d\vk \sqrt{|h|}(\la(\vk)-1)\dot{x}^\mu\frac{\partial p}{\partial\dot{x}^\mu},
\end{equation}
where it is assumed in the approximate equality that $\la(\vk)\rightarrow1$. To put it in another way,
\begin{equation}\label{enthalpy}
    w=-|h|^{-1/2}\rho^a\frac{\de S}{\de\rho^a}=-\dot{x}^\mu\frac{\partial p}{\partial \dot{x}^\mu},
\end{equation}
in the adapted system of coordinates. As a result, the density of the canonical Hamiltonian reads
\begin{equation}\label{Hcan_x_k}
    H_{can}[x(\vk)]=-\sqrt{|q|}\be w=-\sqrt{|q|}\s,
\end{equation}
that is, in the Fock-Taub formalism, the canonical Hamiltonian for the fields $x(\vk)$ coincides with the total entropy taken with an opposite sign. This result can be foreseen from the following considerations. The equations of motion \eqref{hydr_eqm} possesses a conserved charge coinciding with the total entropy of a fluid. Inasmuch as the Lagrangian \eqref{action_FT} does not depend on $\tau$ explicitly, the corresponding canonical Hamiltonian is an integral of motion too. Hence, in a general position, this integral should be proportional to the total entropy. One can also add to the integral of motion a linear combination with constant coefficients of the integrals of motion following from \eqref{vort_cons_law}. However, the only natural constant vector in the model at hand is $\rho^a$ (in the adapted system of coordinates) and we saw that $\rho^af_a$ is proportional to the entropy density.

Formula \eqref{enthalpy} is a generalization of the well-known thermodynamic relation for the canonical ensemble
\begin{equation}
    w=-\be\frac{\partial p}{\partial\be},
\end{equation}
where $p$ is the pressure. Carrying out the standard reasonings analogous to those used for the systems with the Hamiltonians equal to the total energy, one can see that the part of a fluid characterized by bounded values of the coordinates $\s^i$ on the manifold $N$ tends to increase its entropy, i.e., to diminish the magnitude of the canonical Hamiltonian corresponding to this part of fluid. The phonons with the mode functions diagonalizing the quadratic part of the Hamiltonian take away the negative entropy to spatial infinity. This leads to increasing of the entropy of the given part of a fluid. The mention should be made once again that the total entropy of a whole fluid is conserved. Other more effective mechanisms to produce the entropy in a perfect fluid are also possible. For example, it is produced with the help of shock waves or a turbulence (see for details, e.g., \cite{LandLifshHyd}), when the map $x(\vk)$ ceases to be one-to-one. However, this issue needs a further investigation.

The density of a canonical Hamiltonian in the space-time $M$, i.e., for the fields $\vk^a(x)$, coincides with the energy density as it should be. This canonical Hamiltonian is not conserved for the case of the non-stationary matter and gravitational fields. Indeed, in virtue of the general covariance, the action \eqref{action_FT} can be rewritten in the form
\begin{equation}\label{action_FT_inv}
    S[\vk(x)]=\int_M dx\sqrt{|g|(x)}p[\rho^a(\vk(x))e_a^\mu,g_{\mu\nu}(x)],\qquad e^\mu_a=\Big(\frac{\partial\vk^a}{\partial x^\mu}\Big)^{-1},
\end{equation}
i.e., as a functional of the fields $\vk(x)$ defined on the manifold $M$. The equations of motion reads
\begin{equation}
    e^a_\mu\frac{\de S}{\de\vk^a}=-\sqrt{|g|}(\mathcal{L}_\xi S_\mu+\nabla_\nu\xi^\nu S_\mu)=-\sqrt{|g|}\nabla_\nu T^\nu_\mu=0,
\end{equation}
where $S_\mu$ is defined similarly to $\Ga_\mu$. Notice that, in taking a variation with respect to $\vk(x)$, the argument of $\rho^a$ is also varied. If the action functional \eqref{action_FT_inv} does not contain higher derivatives of the fields $\vk(x)$ then, employing the relation,
\begin{equation}
    \frac{\partial e^\mu_b}{\partial\partial_t\vk^a}=-e^\mu_a e^0_b,
\end{equation}
it is easy to show that the density of the canonical Hamiltonian,
\begin{equation}
    H_{can}[\vk(x)]=\sqrt{|g|}\Big(\partial_t\vk^a\frac{\partial p}{\partial\partial_t\vk^a}-p\Big)=-\sqrt{|g|}\Big(\xi^0\frac{\partial p}{\partial\xi^0}+p\Big)=\sqrt{|g|}(wu^0u_0-p)=\sqrt{|g|}T^0_0,
\end{equation}
coincides with the standard expression for the energy density. Here $u^\mu=\xi^\mu/\sqrt{\xi^2}$ is the four-velocity of a fluid flow. Henceforward, we shall consider the Fock-Taub representation of relativistic hydrodynamics in terms of the fields $x(\vk)$ rather than $\vk(x)$ since the symmetry transform \eqref{symm_hydr} looks simpler in this representation. However, it may happen that the Fock-Taub formalism in terms of $\vk(x)$ is more appropriate for some problems.

\section{Quantization of hydrodynamics}\label{QuantHyd}

As we have already noted, according to the background field method \cite{DeWGAQFT,BuchOdinShap}, to construct the quantum field theory, it is sufficient to know the quantum dynamics of small fluctuations on an arbitrary background. The quantum theory of a non-relativistic fluid is also developed in this way \cite{LandLifstat}. Thereupon our task will be to describe the dynamics of small fluctuations (the phonons) of a relativistic fluid taking into account the conditions 1-3 given in Sect. \ref{QGA}. The one-loop calculations performed in \cite{olep,gmse,KalKaz1,KalKaz2} show (see the condition 3) that the action of hydrodynamics \eqref{action_FT} is highly nonlinear. Therefore, we expect that quantum hydrodynamics is perturbatively nonrenormalizable. In order to conduct consistent calculations, we stick to the effective field theory approach (see, e.g., \cite{Weinb,DonGolHol}) supposing that the gradients of the background fields are small, i.e., in fact, considering the infrared limit of a theory.

\subsection{Classical potential}\label{ClassPot}

If we neglect the gradients of all the fields and take into account that the only known macroscopic fields with nonzero vacuum expectation values are the metric $g_{\mu\nu}$, the Higgs field $\phi$, and the vector field $\xi^\mu$ (we also neglect the small, but possibly non-vanishing, contributions depending on the electromagnetic field) then, from the condition 3, the expression for the fluid pressure follows (see, e.g., \cite{KalKaz1})
\begin{equation}\label{eff_pot}
    p_0(\rho^2,|\phi|^2)=\sum_{k=0}^2\sum_{l=0}^\infty a_{kl}(|\phi|^2)t^{2k}\ln^lt^2,
\end{equation}
where $t_a:=\rho_a/\rho^2$. Such a peculiar structure of the classical potential is caused by the fact that the regularization parameter of the energy cutoff regularization always enters the nonrenormalized effective action in the combination $\La^2t^2$, and the general structure of divergencies has the form \eqref{eff_pot} provided the terms suppressed by the Planck mass are omitted (see the remark in Sect. \ref{QGA} after the condition 3). The quantities $a_{kl}$ are the curvature independent gauge invariant scalars with a mass dimension 4,
\begin{equation}
    a_{kl}(|\phi|^2)=\sum_{s=0}^{2-k}b_{kls}|\phi|^{2s},\qquad b_{kls}=const,
\end{equation}
that is, $\rho^a$ is a dimensionless quantity. The logarithmic corrections describe the so-called anomalous scaling (see, e.g., \cite{ColemB}) on the cutoff scale $\La$ and are assumed to be small. The whole expression \eqref{eff_pot} specifies the energy-time anomaly \cite{prop,KalKaz1} of the effective action so long as the gradients of the fields are discarded.

The flat space-time limit corresponds to $t^2=\rho^2=1$. Indeed, for a weak slowly varying gravitational field, the vector field $\xi^\mu$ is close to the Killing vector of the corresponding stationary metric approximating the slowly varying one since in a stationary case the Killing vector field is a solution to the equations of motion \eqref{hydr_eqm} \cite{gmse,FrZel}. So, in that case, we have in the adapted system of coordinates (see, e.g., \cite{LandLifshCTF})
\begin{equation}\label{poten_New}
    \xi^2=\rho^2=e^{2\vf_N}\approx1+2\vf_N,
\end{equation}
where $\vf_N$ is the Newtonian potential. Of course, one can normalize the vector field $\rho^a$ to any other positive constant in a flat space-time. In a certain sense, this corresponds to a shift of the Newtonian potential by a constant. Furthermore, it is natural to redefine the field as $\rho^a\rightarrow\La^{-1}_{dm}\rho^a$, where $\La_{dm}$ is some constant parameter with a unit mass dimension, such that $\rho^a$ has the mass dimension 1. Then the entropy density of a fluid \eqref{temp_entrop} will possess the standard mass dimension 3. It has the dimension of an energy density for the normalization condition $\rho^2=1$ in a flat space-time. However, from an utilitarian point of view, these dilatations of the vector $\rho^a$ just results in the redefinition of the constants $a_{kl}$ and does not change the observables. Such transformations resemble the renormalization group transformations and coincide with them under the condition that the coefficients at the logarithms in \eqref{eff_pot} coincide with the scaling functions $\be$ and $\ga$. Thus we put $\rho^2=1$ in a flat space-time.

The classical pressure \eqref{eff_pot} should satisfy the conditions 1 and 2 in the leading order of the perturbation theory. Since
\begin{equation}
    T^{ab}_0=-2p_0'\rho^a\rho^b-p_0h^{ab}=w_0u^au^b-p_0h^{ab},
\end{equation}
where the prime denotes the derivative with respect to $\rho^2$ and $u^a:=\rho^a/\sqrt{\rho^2}$, then
\begin{equation}\label{cond_tem_zero}
    \left.p_0(\rho^2,|\phi|^2)\right|_{|\phi|=\eta,\rho^2=1}=0,\qquad \left.p'_0(\rho^2,|\phi|^2)\right|_{|\phi|=\eta,\rho^2=1}=0,
\end{equation}
where $\eta\approx 247$ GeV is the vacuum expectation value of the Higgs field in a flat space-time. Is also supposed that the term $a_{00}$ in $p_0$ contains the classical potential of the Higgs field in a flat space-time limit. Therefore, the following relations should hold
\begin{equation}\label{cond_ewmin}
    \left.\dot{p}_0(\rho^2,|\phi|^2)\right|_{|\phi|=\eta,\rho^2=1}=0,\qquad 4\eta^2\left.\ddot{p}_0(\rho^2,|\phi|^2)\right|_{|\phi|=\eta,\rho^2=1}=-m^2_\chi,
\end{equation}
where the dot denotes the derivative with respect to $|\phi|^2$ and $m_\chi\approx125$ GeV is the Higgs boson mass in a flat space-time. The normalization conditions \eqref{cond_tem_zero}, \eqref{cond_ewmin} are only necessary ones. The experimental data may lead to additional constraints on $p_0$. For instance, it was shown in \cite{gmse} that, in order to reach an agree of the theory with the experimental data concerning the gravitational red-shift effect, one has to demand
\begin{equation}\label{cond_redshift}
    \left.\dot{p}'_0(\rho^2,|\phi|^2)\right|_{|\phi|=\eta,\rho^2=1}=0.
\end{equation}
Let us stress once again that the normalization conditions \eqref{cond_tem_zero}, \eqref{cond_ewmin}, and \eqref{cond_redshift} are imposed on the effective action and are valid for the classical potential only in the leading order of the perturbation theory. Having calculated the quantum corrections to the classical potential, we need to impose this conditions to fix the arbitrary constants appearing in the effective action. It is the effective action which is observed in an experiment.

\subsection{Polytropic equation of state}\label{PolEoS}

In this paper, we identify the fluid resulting from the existence of the quantum gravitational anomaly with a dark matter or its considerable part. In the limit of a weak slowly varying gravitational field, it follows from \eqref{cond_tem_zero} that $w_0\approx\e_0\approx\s_0\sim x$ and $p_0\sim x^2$, where, for short, we have introduced the notation  $x:=1-\rho^2$ and $\e_0$ is the energy density of a fluid. In a general case, we can adjust the constants in $p_0$ in such a way that the first $n$ derivatives of $p_0$ with respect to $\rho^2$ at $\rho^2=1$ vanish. Then we have
\begin{equation}\label{asympt_wfl}
    \e_0\approx w_0=-2\be^2p_0'\approx2Ax^n,\qquad p_0\approx A\frac{x^{n+1}}{n+1},\quad n\in\mathbb{N},
\end{equation}
where $A$ is a positive constant with the mass dimension 4. Rearranging these expressions, we arrive at
\begin{equation}\label{eos_poly}
    p_0=\frac{\e_0^{1+1/n}}{2(n+1)(2A)^{1/n}}=:K\e_0^{1+1/n}=\frac{\e_0x}{2(n+1)}.
\end{equation}
This is the equation of a polytrope (see, e.g., \cite{Hored}) with the index $n$ and the polytropic constant $K$. If we identify $\La_{dm}x/[2(n+1)]$ with $kT$ then the equation of state \eqref{eos_poly} coincides with the equation of state for a perfect gas. The limit $n\rightarrow\infty$ corresponds to the isothermal polytropic process. The speed of sound in the weak field limit becomes \cite{gmse,prop,KalKaz1}
\begin{equation}\label{sound_speed}
    c_s^{2}=\frac{\partial p_0}{\partial\e_0}=-\frac{p_0'}{3p_0'+2\beta^2p_0''}\approx\frac{x}{2n}.
\end{equation}
Thus, in the weak field limit, the relativistic fluid stemming from the existence of the quantum gravitational anomaly behaves as a perfect gas in a polytropic process with a universal polytropic constant $K$ and a natural polytropic index $n$. This fact can be used to falsify the hypothesis about the identification of this fluid with a dark matter.

Notice that, in a stationary case, when the fluid is in a hydrostatic equilibrium, the asymptotics \eqref{poten_New}, \eqref{asympt_wfl} imply that this fluid is absent in a flat space-time limit. This guarantees the absence of a Lorentz-invariance violation as we have demanded imposing the normalization conditions. On the other hand, as long as the fluid itself produces the gravitational field, there may exist stable self-sustained configurations consisting of this fluid and the gravitational field only. The cosmological constant and the cosmological expansion caused by it decrease the stability of such gravitationally bounded systems \cite{BAMoNo,ChernUFN}. So these stable self-sustained configurations may exist only if their density is larger than a certain critical one \cite{BAMoNo,MerBisTar}.

We can estimate the value of the constant $A$ and, consequently, of the constant $K$ assuming that the velocity of motion of the dark matter with respect to the Galaxy center in the Sun neighbourhood is non-relativistic. In this case, we can set $x\approx-2\vf_\odot$ and take \cite{GarLiReLa,Read,ZRVLZ,KSLBN} (see, however, \cite{BCMS})
\begin{equation}\label{dm_sun}
    |\vf_\odot|\sim 10^{-6},\qquad\e_{\odot}\sim 0.5\, \text{GeV}/\text{cm}^3,
\end{equation}
whence
\begin{equation}\label{A_const}
    A=\frac{\e_\odot}{2|2\vf_\odot|^n}.
\end{equation}
The concrete values of $A$ for several different $n$ are as follows
\begin{equation}\label{A_const_vls}
\begin{aligned}
    A_1&\approx1.25\times10^5\,\text{GeV}/\text{cm}^3\approx(1\,\text{eV})^4,&\quad A_4&\approx1.6\times10^{22}\,\text{GeV}/\text{cm}^3\approx(19\,\text{keV})^4,\\ 
    A_8&\approx9.8\times10^{44}\,\text{GeV}/\text{cm}^3\approx(9.3\,\text{GeV})^4\approx(\eta/26)^4,&\quad A_9&\approx4.9\times10^{50}\,\text{GeV}/\text{cm}^3\approx(248\,\text{GeV})^4\approx\eta^4.
\end{aligned}
\end{equation}
At first sight, it appears that the natural value of $n$ for the electroweak scale is $n=9$. However, as we shall see, such a value is $n=8$ since a small factor of the order of $1/n!$ diminishing the magnitude of the constant $A$ arises in developing the pressure $p_0$ as a Taylor series in $x$. For $n<8$, the constants in the coefficients $a_{kl}(|\phi|^2)$ have to be adjusted in such a way as to diminish the energy density and, consequently, the magnitude of the constant $A$ from the natural electroweak scale to an acceptable value.

The models of a dark matter with a polytropic equation of state have been already considered in the literature. As well-known, the conventional Navarro-Frenk-White (NFW) profile \cite{NaFrWhit} for the dark matter density in galaxies possesses a cusp at the galaxy center which is not observed in the astrophysical data. The observed profiles are more likely described by the so-called cored distributions (for their scaling properties, see, e.g., \cite{DGSFMWGGKW}). The NFW profile fits the results of the $N$ body simulations and the cusp in it is due to a pressureless of the dark matter constituted by weakly interacting massive particles (the dust). The polytropic equation of state introduces a non-vanishing pressure and so avoids the cusp problem. In the papers \cite{SaxFer,Saxt,SaxWu,SaxSorWu}, a dark matter with self-interaction was studied. Its equation of state was taken to be \eqref{eos_poly} with an arbitrary $n$ and the interval $3.5\leq n\leq5$ was approved as the most probable one \cite{SaxWu,SaxSorWu,SaxFer}. The polytropic equation of state also arose in employing the Tsallis formalism to describe the dark matter. The latter was supposed to consist of collisionless particles in a non-equilibrium thermodynamical state (see \cite{CRMNSZ,CFSBSRG} and the critique in \cite{FerHjo}). These papers suggested the optimal value of the polytropic index to be $n\approx5$. In the paper \cite{BAMoNo}, it was argued that, out of the cusp, the NFW profile for the dark matter density can be obtained by using the polytropic fluid with the index $n=3$. In describing a dark matter as a Bose-Einstein condensate of some scalar field (see, e.g., \cite{Peebl,Goodm,HarMoc,DwKerGer}), the polytropic index is found to be $n=1$ in the non-relativistic limit.

The cosmological data can also impose certain constraints on the parameters of the equation of state \eqref{eos_poly}. We shall not give here a detailed analysis of the cosmological model with a dark matter of the form \eqref{eff_pot}, \eqref{eos_poly}, but only check that the fluid with a polytropic equation of state does provide a model of a cold dark matter. Let us pass to the system of coordinates where the fluid at issue is at rest in average and its density is homogeneous, i.e.,
\begin{equation}
    \xi^\mu=(\xi^0(t),0,0,0)\quad\text{and}\quad ds^2=dt^2-a^2(t)\ga_{ij}dx^idx^j,
\end{equation}
where $ds^2$ is the interval squared of the Friedmann-Lema\^{\i}tre metric. At the present stage of the Universe evolution, the average density of the dark matter is small \cite{GorbRub1,Schneidr}
\begin{equation}\label{dm_now}
    \e_{dm}=\Omega_{dm}\e_c\sim10^{-6}\,\text{GeV}/\text{cm}^3,\qquad\e_c=0.53\times10^{-5}\,\text{GeV}/\text{cm}^3,
\end{equation}
where $\e_c$ is the critical energy density and $\Omega_{dm}=0.20$ is the fraction of a dark matter in the energy content of the Universe. Therefore, we can consider that the fluid is in a weak field regime $\xi^2\approx1$ and obeys the equation of state \eqref{eos_poly}. Then the condition of covariant divergenceless of the energy-momentum tensor \eqref{hydr_eqm} possesses the solution
\begin{equation}\label{dm_sol_FRW}
    \e=\e_{dm}\Big(\frac{a_0}{a}\Big)^{3}\Big\{1+K\e_{dm}^{1/n}\Big[1-\Big(\frac{a_0}{a}\Big)^{3/n}\Big]\Big\}^{-n}\approx2Ax^{n},
\end{equation}
where $a_0$ is the present-day value of the scale factor. This solution corresponds to a cold dark matter (a dust) so long as $K\e_{dm}^{1/n}\ll1$. From \eqref{eos_poly} and \eqref{A_const}, we obtain
\begin{equation}
    K=\frac{(2A)^{-1/n}}{2(n+1)}=\frac{|\vf_\odot|\e^{-1/n}_\odot}{n+1},\qquad K\e_{dm}^{1/n}=\frac{|\vf_\odot|}{n+1}\Big(\frac{\e_{dm}}{\e_\odot}\Big)^{1/n}.
\end{equation}
Taking into account the magnitudes \eqref{dm_sun}, \eqref{dm_now} of the quantities entering the last expression, it is evident that $K\e_{dm}^{1/n}\ll10^{-6}$ for any natural number $n$.

The solution \eqref{dm_sol_FRW} allows us to find the cosmological value of $\xi^2(t)$, i.e., the value of $\xi^2$ averaged over the homogeneity cell of the Universe. Neglecting $K\e_{dm}^{1/n}$ in \eqref{dm_sol_FRW}, we come to
\begin{equation}
    x\approx\Big(\frac{\e_{dm}}{2A}\Big)^{1/n}\Big(\frac{a_0}{a}\Big)^{3/n}=2(n+1)K\e_{dm}^{1/n}\Big(\frac{a_0}{a}\Big)^{3/n}= 2|\vf_\odot|\Big(\frac{\e_{dm}}{\e_\odot}\Big)^{1/n}\Big(\frac{a_0}{a}\Big)^{3/n}.
\end{equation}
One can naively surmise that $\ln\xi^2=\ln(a/a_0)$ (compare, e.g., \cite{DeWQFTcspt,olep,gmse,KalKaz1} with \cite{LeonWood,GMPW,WanWood,MarPrad}), but now we see this is not the case.
The solutions of the Friedmann equations for a dark matter with a polytropic equation of state for an arbitrary $n>0$ can be found, for instance, in \cite{Barrow,Odints,Chavan}. The polytropic gas with the negative index $n$ was used, for example, in \cite{KleidSpy,Barrow,Odints} to describe a dark energy. It should be stressed in this connection that, for the model \eqref{eff_pot} we consider, the polytropic equation of state \eqref{asympt_wfl}, \eqref{eos_poly} arises only in the weak field limit. Formulas \eqref{asympt_wfl}, \eqref{eos_poly} are not applicable at the early stages of the Universe evolution.

Notice also another one feature of the equation of state \eqref{asympt_wfl}, \eqref{eos_poly}. For $n$ odd, the energy density of a fluid can be negative, if there are domains in space where $x<0$, i.e., $\rho^2>1$. For a spherically-symmetric case, the problem of a hydrostatic equilibrium of a polytropic fluid is described by the Lane-Emden equation (see, e.g., \cite{Hored}), and it is well-known that such regions exist with a necessity at $n<5$. Usually, the transition from the region with $x>0$ to the region with $x<0$, in increasing the radius $r$, is interpreted as the presence of a sharp boundary of a gas cloud. The part of the solution to the Lane-Emden equation with $r>r_0$, where $r_0$ is the transition point, is thrown away as unphysical: the particle number density is negative at $r>r_0$. This procedure is quite reasonable for a usual gas of particles, but, in our case, it would be unnatural to impose the additional constraint $\e\geq0$. Usually, the domains with a negative energy density appear in the solutions to the Lane-Emden equation at sufficiently large values of $r$. One should also take into account that, at large $r$, the negative contribution of the cosmological constant to a gravitating ``mass'' becomes considerable \cite{ChernUFN,ChernJETP,MerBisTar} since $M_{\La_c}(r)=-8\pi\La_c r^3/3$, where $\La_c$ is the cosmological constant. In particular, for the Local group of galaxies, the contribution of the cosmological constant dominates at $r\geq 1.4$ Mpc \cite{ChernUFN,ChernJETP}. One has to take into account the cosmological expansion and the galaxy formation processes on these scales \cite{LukRub}.

\subsection{Gradient corrections}\label{GradCorr}

In what follows, we shall consider the general case \eqref{eff_pot} with the asymptotics \eqref{asympt_wfl}, \eqref{eos_poly}. In the stationary limit, the gradient corrections to the pressure \eqref{eff_pot} were found in \cite{KalKaz1} to the one-loop approximation.  In this paper, we restrict our study to the leading gradient corrections that are of the second order in derivatives and the homogeneity degree $-2$ under the dilatations $\rho^a\rightarrow\la\rho^a$. One could hope that some structures do not appear in calculating the quantum corrections (this does happen for higher coefficients of the heat kernel expansion, see \cite{Ven}). However, all the possible structures permitted by the two restrictions above arise in our case.

Indeed, in \cite{KalKaz1}, the quadratic divergencies of the second order in derivatives were found in the one-loop effective action of quantum gravity induced by a scalar field. Up to the total derivatives, they are
\begin{equation}
    f^2,\qquad Rt^2,
\end{equation}
where $f_{ab}:=\partial_{[a}t_{b]}$. The one-loop contribution of fermions produces the additional structure \cite{Fursaev1,Fursaev2}
\begin{equation}
    a^2t^2,
\end{equation}
where $a_a:=\partial_a\ln\sqrt{\rho^2}$. These expressions were obtained under the assumption of stationarity of the space-time, that is, the vector $\rho^a$ is the Killing vector for the metric $h_{ab}$ here. In a general case, we have eight independent structures submitting to the restrictions pointed above such that
\begin{equation}\label{p2}
    p_{2}=b_1Rt^2+b_2R_{ab}t^at^b+b_3\nabla_at_b\nabla^at^b+b_4(\nabla_at^a)^2+b_5a^2t^2+b_6(a_at^a)^2+b_7\rho^a\rho^b\nabla_{ab}t^4+b_8\rho^a\nabla_at^4\nabla_b\rho^b.
\end{equation}
All the constants $b_i$ have the mass dimension 2. One can verify that, in the case of a stationary metric with the Killing vector $\rho^a$, all these structures reduce to the three structures presented above up to the total derivatives. Obviously, we can choose other basis of independent structures. However, the basis \eqref{p2} seems to be the most convenient for a subsequent derivation and analysis of the equations of motion.

Our task is to find the constraints on the coefficients $b_i$ that provide
\begin{enumerate}[i)]
  \item The absence of unstable modes (the ghosts) in the weak field limit;
  \item The absence of small perturbations (the phonons) in the flat space-time limit $\rho^2=1$, i.e., in this limit, they must possess the dispersion law of the form $\omega=0$ or $\omega\rightarrow\infty$, where $\omega$ is the energy of the phonon mode.
\end{enumerate}
The necessity of the first condition is evident. The second requirement is needed for the Lorentz-invariance to be preserved in the flat space-time limit, i.e., the condition 1 of Sect. \ref{QGA}. Indeed, so long as the phonons are bosons and can constitute a condensate, the existence of nontrivial phonon branches of the dispersion law inevitably leads to the possibility of macroscopic perturbations of the field $\xi^\mu$ in the flat space-time limit. Consequently, it leads to a violation of the Lorentz-invariance.

Note that the analysis we are about to carry out differs from that is given in the papers \cite{Bhatt1,Bhatt2,Bhatt3}. In those papers, the general equations of motion of a relativistic fluid with viscosity were considered and the conditions were found when the divergence of the entropy current is nonnegative on the solutions to the equations of motion. In our case, the equations of relativistic hydrodynamics are Lagrangian, describe a perfect fluid, and the divergence of the entropy current is zero \eqref{hydr_eqm}.

In order to obtain the contributions of the terms \eqref{p2} to the equations of motion \eqref{hydr_eqm_0}, we need to find the variation of the action \eqref{action_FT} with the pressure \eqref{p2}. It is not difficult to verify that
\begin{equation}\label{variations}
\begin{aligned}
    \de\rho^a&=0,\qquad\de h_{ab}=\nabla_{(a}\theta_{b)},\qquad\de\sqrt{|h|}=\sqrt{|h|}\nabla_a\theta^a,&\quad\de a_a&=\nabla_a(t^b\rho^c\nabla_b\theta_c),\\
    \de\Ga^c_{ab}&=\tfrac12(\nabla_{(ab)}\theta^c-R^c_{\ (ab)k}\theta^k)=\nabla_{ba}\theta^c+R^c_{\ akb}\theta^k,&\quad\de t^a&=-2t^at^b\rho^c\nabla_b\theta_c,\\
    \de t^{2k}&=-2kt^{2k}t^a\rho^b\nabla_a\theta_b,&\de(\nabla_a\ups^a)&=\ups^a\nabla_{ab}\theta^b+\nabla_a\de\ups^a,
\end{aligned}
\end{equation}
where $\theta^a:=e^a_\mu\de x^\mu$. Therefore, the variation with respect to $x^\mu$ is equivalent to the evaluation of the Lie derivative along the vector field $\theta^a$ with the only difference that the variation of $\rho^a$ vanishes. Formally, one may put $\mathcal{L}_\theta\rho^a=0$. Taking the variation of the terms \eqref{p2} in the action \eqref{action_FT}, we deduce
\begin{equation}\label{eqmot_18}
\begin{aligned}
    E^1_c&=2\nabla_a(t^at_cR)-R\nabla_ct^2,\qquad\qquad\qquad\qquad\qquad\quad E^2_c=\nabla_d(4R_{ab}t^at^bt^d\rho_c-2R_{ca}t^at^d)-R_{ab}\nabla_c(t^at^b),\\
    E^3_c&=2\big[\nabla^2t_a\nabla_ct^a+\nabla_a(t^a\nabla^2t_c-2t^at^b\rho_c\nabla^2t_b)\big],\qquad E^4_c=\nabla_c\big[(\nabla_at^a)^2+2t^a\nabla_{ab}t^b\big]-4\nabla_d(t^dt^a\rho_c\nabla_{ab}t^b),\\
    E^5_c&=2\big[\nabla^2t\nabla_ct+\nabla_a(t^au_c\nabla^2t)\big],\qquad\qquad\qquad\quad
    E^6_c=2\nabla_d\big\{t^d\rho_c[\nabla_b(t^bt^aa_a)+2(a_at^a)^2]\big\}-\nabla_c\big[(a_at^a)^2\big],\\
    E^7_c&=4\nabla_d\big[t^2t^dt_c\nabla_{ab}(t^at^b)\big]-\nabla_c(\rho^a\rho^b\nabla_{ab}t^4)-\nabla_{ab}(\rho^a\rho^b\nabla_ct^4)-\rho^a\rho^b\nabla_dt^4R^d_{\ acb},\\
    E^8_c&=\nabla_c\big[\rho^b\nabla_b(\rho^a\nabla_at^4)\big]-4\nabla_d\big[t^2t^dt_c\nabla_a(\rho^a\nabla_b\rho^b)\big],
\end{aligned}
\end{equation}
where $t:=\sqrt{t^2}$ and the index at $E_c$ marks the number of the term in \eqref{p2}. The term without derivatives \eqref{eff_pot} gives
\begin{equation}\label{eqmot_0}
    E^0_c=-\nabla_cp_0-2\nabla_a(\rho^a\rho_cp_0').
\end{equation}

\subsection{Linearized equations of motion}\label{LinEqMot}

The background field method is realized in the standard way. The field $x(\vk)$ is represented in the form
\begin{equation}
    x^\mu(\vk)=:\bar{x}^\mu(\vk)+\theta^\mu(\vk)=:\bar{x}^\mu+\bar{e}^\mu_a\theta^a,
\end{equation}
where $\bar{x}(\vk)$ is an arbitrary background field that afterward will be identified with the vacuum expectation value of the operator $x(\vk)$, and $\bar{e}^\mu_a=\partial_a\bar{x}^\mu$. It is assumed that the vacuum state of a fluid is the ground state of the Hamilton operator associated with \eqref{Hcan_x_k}. Since the Hamilton operator does not depend on the time $\tau$ explicitly, its ground state evolves in a trivial manner when the map $\bar{x}(\vk)$ is one-to-one. So one can use the in-out effective action to calculate the vacuum averages of operators. From the quantum field theory formalism viewpoint, the entropy can be generated by the non-perturbative processes only when the map $x(\vk)$ develops the fold singularities.

The symmetry transform \eqref{symm_hydr} acts as \cite{DeWGAQFT,BuchOdinShap}
\begin{equation}
    \de_\e\bar{x}^\mu(\vk)=0,\qquad\de_\e\theta^a=\e^a+\e^b\tilde{\nabla}_b\theta^a,\qquad (\tilde{\nabla}_a)^b_c:=\partial_a\de^b_c+\bar{e}^b_\nu\partial_a\bar{e}^\nu_c,
\end{equation}
where $\tilde{\nabla}_a$ is a trivial symmetric connection constructed with the help of the Jacobi matrices $\bar{e}_a^\mu$. The kernel of the symmetry generator looks as follows
\begin{equation}
    \de_\e\theta^B=\e^\al R_\al^B,\qquad R_\al^B=(\de^{b}_a+\tilde{\nabla}_a\theta^b(\vk'))\de(\s-\s'),
\end{equation}
where $\al:=(a,\s)$ and $B:=(b,\vk')$. Obviously, these generators form an algebra
\begin{equation}
    [\de_{\e_1},\de_{\e_2}]=-\de_{[\e_1,\e_2]}.
\end{equation}
The effective action of the quantum fields $\theta^a$ depends parametrically on the background fields $\bar{x}(\vk)$. It is also invariant with respect to the transformations \eqref{symm_hydr}. This symmetry of the effective action can be violated only by the quantum anomalies. However, the physical regularization, that we imply, is covariant under the transformations \eqref{symm_hydr} and cannot destroy this symmetry. Also there is not any additional structures in the effective action as, for example, in the case of a scalar field on a curved background (see, for details, \cite{GriMaMos,KalKaz1,gmse,olep,Page,BrOtPa,FrZel,FrolNov,AndHisSam}). The fluid Hamilton operator defining the dynamics and the vacuum state is expressed in terms of the fields $\rho^a$ and $x(\vk)$ which are already present in the effective action. Therefore, we have the Ward identities
\begin{equation}
    R_\al^A\frac{\de\Ga}{\de\theta^A}\equiv0,\qquad R_\al^A\frac{\de^2\Ga}{\de\theta^A\de\theta^B}+\frac{\de R_\al^A}{\de\theta^B}\frac{\de\Ga}{\de\theta^A}\equiv0,\qquad R_\al^A\frac{\de^3\Ga}{\de\theta^A\de\theta^B\de\theta^C}+\frac{\de R_\al^A}{\de\theta^{(B}}\frac{\de^2\Ga}{\de\theta^A\de\theta^{C)}}\equiv0, \quad\text{etc.}
\end{equation}
In particular, the relation for the polarization operator follows from the second identity
\begin{equation}
    \int d\tau\Pi^{ab}(\tau,\s;\tau'\s')=0,\qquad \Pi^{AB}:=(D^{-1})^{AB}-\frac{\de^2\Ga}{\de\theta^A\de\theta^B}\Big|_{\theta^A=0},
\end{equation}
on the solutions to the quantum equations of motion. Here $D^{ab}(\vk;\vk')$ is the propagator of the fields $\theta^a$ on the background $\bar{x}(\vk)$. In fact, the Ward identities found express the entropy and vorticity potential conservation laws.

In order to investigate the dynamics of small perturbations of the field $x(\vk)$, which are determined by the classical equations of motion, it is necessary to linearize the expressions \eqref{eqmot_18} and \eqref{eqmot_0} over the background $\bar{x}(\vk)$ (henceforward the bar is omitted). To this aim, it is useful to employ formulas \eqref{variations} once again. The expressions for the linearized equations of motion are rather huge and given in Appendix \ref{LinEQM}. We only write here the linearized expression \eqref{eqmot_0}:
\begin{multline}\label{E0_lin}
    \de E^0_c=-2\big[\nabla_c(\rho^a\rho^bp_0'\nabla_a\theta_b)+\nabla_a(\rho^a\rho^bp_0')\nabla_{(c}\theta_{b)}+\rho^a\rho^bp_0'(\nabla_{ab}\theta_c+\theta^sR_{sbca})+\\
    +\rho^a\rho_cp_0'\nabla_{ab}\theta^b+2\nabla_a(\rho^a\rho_c\rho^b\rho^sp_0''\nabla_b\theta_s) \big].
\end{multline}
This expression will be needed for us in deriving the quadratic in $\theta^a$ terms in the fluid equations of motion.

In the limit when the gradients of the metric field can be neglected and
\begin{equation}\label{flat_lim}
    \rho^a=\be\de^a_0,\qquad t^a=\be^{-1}\de^a_0,\qquad\be=\sqrt{\rho^2}=const,
\end{equation}
it follows from formulas \eqref{E0_lin} and \eqref{lin_eqm_ap} that
\begin{equation}\label{eqm_sec_ord}
\begin{gathered}
  \de E^0_c=-2\be^2p_0'\Big[\partial_c\dot{\theta}_0+\ddot{\theta}_c+\de^0_c\Big(\partial_a\dot{\theta}^a+2\be^2\frac{p_0''}{p_0'}\ddot{\theta}_0\Big)\Big],\qquad \de E^1_c=0,\qquad \de E^2_c=0,\qquad\de E^3_c=2\be^{-2}\Box\ddot{\theta}_c,\\
  \de E^4_c=2\be^{-2}\big[\partial_{ca}\ddot{\theta}^a-2\partial_c\dddot{\theta}_{\hspace{-0.11em}0}-2\de_c^0(\partial_a\dddot{\theta}^a-2\overset{(4)}{\theta}_{\hspace{-0.17em}0} ) \big],\qquad
  \de E^5_c=2\be^{-2}\de_c^0\Box\ddot{\theta}_0,\qquad \de E^6_c=2\be^{-2}\de^0_c\overset{(4)}{\theta}_{\hspace{-0.17em}0},\\ \de E^7_c=8\be^{-2}(\partial_c\dddot{\theta}_{\hspace{-0.11em}0}+\de^0_c\partial_a\dddot{\theta}^a),\qquad \de E^8_c=-4\be^{-2}(\partial_c\dddot{\theta}_{\hspace{-0.11em}0}+\de^0_c\partial_a\dddot{\theta}^a),
\end{gathered}
\end{equation}
where $\Box:=h^{ab}\partial_a\partial_b$ is the d'Alembert operator and the dots denote the derivatives with respect to $\tau$. The first and second expressions vanish not only in the limit referred above, but for any vacuum solution to the Einstein equations $R_{ab}=0$. The last two expression are similar in the limit \eqref{flat_lim}.

Let us consider, at first, the linearized equations of motion \eqref{E0_lin} in the limit \eqref{flat_lim}. In this case,
\begin{equation}\label{eqm_lin_secord}
    \de E^0_i=-2\be^2p_0'(\partial_i\dot{\theta}_0+\ddot{\theta}_i)=0,\qquad\de E_0^0=2\be^2p_0'(c_s^{-2}\ddot{\theta}_{\hspace{-0.11em}0}+\partial_i\dot{\theta}_i)=0.
\end{equation}
It is convenient to split the perturbation $\theta_i$ into the longitudinal and transverse components
\begin{equation}\label{decomp_pot_vor}
    \theta_i=\theta^\parallel_i+\theta^\perp_i,\qquad\theta^\parallel_i=\partial_i\theta,\qquad \partial_i\theta^\perp_i=0.
\end{equation}
Then the longitudinal modes obey the standard wave equation describing the propagation of phonons
\begin{equation}\label{prll_mod}
    \dot{\theta}_0=-\ddot{\theta},\qquad\de E_0^0=-2\be^2p_0'(c_s^{-2}\dddot{\theta}-\Delta\dot{\theta})=0,
\end{equation}
where we have assumed the zero boundary conditions at spatial infinity. The dispersion law takes the form $\omega^2=c^2_s\spk^2$ and reduces to $\omega^2=0$ in the flat space-time limit. The dynamics of transverse modes are described by
\begin{equation}
    \dot{\theta}_0=0,\qquad\ddot{\theta}^\perp_i=0,\qquad\de E_0^0=2\be^2p_0'c_s^{-2}\ddot{\theta}_{0}=0.
\end{equation}
This is the so-called entropy-vortex perturbations (see, e.g, \cite{LandLifshHyd}). They possess a trivial dispersion law $\omega^2=0$, that is, they do not depend on the Lagrangian time $\tau$ and move along with the fluid flow. The existence of such peculiar perturbations is a consequence of the vorticity conservation law \eqref{vort_cons_law}.

These entropy-vortex perturbations represent a certain problem for a straightforward application of the standard quantization procedure \cite{ZhuKlau,ENRW,DHNS,Torri,BallBell,GripSuth,Ballest} -- the propagator does not decrease for large $|\spk|$ that results in severe ultraviolet divergencies, and it possesses a singularity at $\omega=0$ that leads to the presence of infrared divergencies. The physical solution of this problem, in contrast to the formal one \cite{ZhuKlau}, consists in that one needs to take into account the gradient corrections to the pressure and nonlinear terms in the equations of motion. These corrections modify the dispersion law.

The gradient terms of the second order in derivatives \eqref{p2} do change the dispersion law for the transverse modes. However, we shall see now that this modification is prohibited by the normalization condition (ii). Multiplying the contributions \eqref{eqm_sec_ord} by the corresponding coefficients $b_i$ and add them, we derive
\begin{equation}\label{eqm_lin_fl}
\begin{split}
  2\be^{-2}&\Big[(b_3+b_4+b_5+b_6+8b_7-4b_8)\overset{(4)}{\theta}_{\hspace{-0.17em}0}-(b_3+b_5)\Delta\ddot{\theta}_0+(b_4-4b_7+2b_8)\partial_i\dddot{\theta}_{\hspace{-0.11em}i} +\beta^4p_0'(c_s^{-2}\ddot{\theta}_0+\partial_i\dot{\theta}_i)\Big]=0,\\
  2\be^{-2}&\Big[b_3(\overset{(4)}{\theta}_{\hspace{-0.17em}i}-\Delta\ddot{\theta}_i)-b_4\partial_{ij}\ddot{\theta}_j-(b_4-4b_7+3b_8)\partial_i\dddot{\theta}_{\hspace{-0.11em}0} -\beta^4p_0'(\partial_i\dot{\theta}_0+\ddot{\theta}_i) \Big]=0.
\end{split}
\end{equation}
For the transverse modes of the form
\begin{equation}
    \dot{\theta}_a=(0, \dot{\theta}^\perp_i),
\end{equation}
we arrive at
\begin{equation}
    b_3\Box\ddot{\theta}^\perp_i-\beta^4p_0'\ddot{\theta}^\perp_i=0.
\end{equation}
Whence the dispersion law follows
\begin{equation}
    \omega^2=0,\quad\text{or}\quad \omega^2=-\beta^4p_0'/b_3+\spk^2.
\end{equation}
In the latter case we have the dispersion law of a relativistic massive particle with a mass squared $-\beta^4p_0'/b_3$. Bear in mind the analogy with the sound waves in an isotropic elastic medium (see, e.g., \cite{LandLifshElas}), one may expect the existence of such a dispersion law for transverse modes when the gradient corrections \eqref{p2} are taken into account. Similar gradient corrections to the free energy of a solid result in the nontrivial dispersion law for the transverse perturbations. In the flat space-time limit, this branch of the dispersion law describes the massless perturbations and does not die out. This contradicts the condition (ii). Hence, we set
\begin{equation}\label{b3}
    b_3=0,
\end{equation}
and the issue of the transverse modes is left. We shall return to this problem in the next section regarding the nonlinear contributions to the equations of motion.

\subsection{Stability criteria}\label{StabCrit}

Let $\e_a(\omega,\spk)$ be the eigenvalues of the Fourier transform of the matrix operator \eqref{eqm_lin_fl} acting on $\theta^a$. Then the dispersion law determined by the equation
\begin{equation}\label{disp_law_det}
    \e_a(\omega_a(\spk),\spk)=0,\quad\forall\spk,
\end{equation}
describes stable perturbations as long as \cite{Migdal}
\begin{equation}\label{stabil_crit}
    \omega_a(\spk)\in \mathbb{R},\qquad\omega\e_a'(\omega)\Big|_{\omega=\omega_a(\spk)}>0,\quad\forall\spk.
\end{equation}
It is useful to rewrite the latter condition as
\begin{equation}\label{stabil_crit2}
    \frac{\partial\e_a}{\partial\omega^2}\Big|_{\omega=\omega_a(\spk)}>0.
\end{equation}
It ensues the correct sign of the propagator residue at the pole in the $\omega$ plane: it is positive for $\omega>0$ and negative for $\omega<0$. The reality condition for $\omega_a(\spk)$ can be weaken to some extent to allow for unstable particles. Then it is required that
\begin{equation}\label{stabil_crit_gen}
    \re\omega_a(\spk)\im\omega_a(\spk)\leq0\;\text{and}\;\re\omega\re\e_a'(\omega)\Big|_{\omega=\omega_a(\spk)}>0.
\end{equation}
However, this case is not realized for the wave operators corresponding to the Lagrangian equations of motion, i.e., to the equations of motion coming from a variational principle.

In the framework of the effective field theories, i.e., the derivative expansion that we are employing now, the terms with higher derivatives with respect to time (of the third order and higher) in the equations of motion cannot produce extra branches in the dispersion law provided the conditions \eqref{stabil_crit} are fulfilled. These terms just modify, at large energies and momenta, the dispersion law given by the equations of motion of the second order in derivatives with respect to time \eqref{eqm_lin_secord}. Indeed, expanding the action in derivatives, we obtain from it the wave operator polynomial in $\omega$ and $\spk$. This operator possesses the bounded eigenvalues $\e_a(\omega,\spk)$ at finite $\omega$ and $\spk$. Equation \eqref{disp_law_det} has two roots, at least, for small $\omega$ and $\spk$. These two roots satisfy \eqref{stabil_crit} and, up to small corrections, are determined by the theory without higher derivatives \eqref{eqm_lin_secord}. Under sufficiently weak restrictions on the wave operator of the Lagrangian equations of motion, the eigenvalues $\e_a(\omega)$ obey the Schwarz symmetry principle (see, e.g., \cite{Newton_scat})
\begin{equation}
    \e_a^*(\omega)=\e_a(\omega^*).
\end{equation}
So, if \eqref{disp_law_det} possesses a complex solution then a complex conjugate to it is also a solution to \eqref{disp_law_det}. Hence, for a stable model, equation \eqref{disp_law_det} has only two nondegenerate roots (referred above) in the complex $\omega$ plane. Otherwise, as long as \eqref{disp_law_det} has some additional complex roots, they necessarily violate the condition \eqref{stabil_crit_gen}. If these extra roots are real then the continuity of $\e_a(\omega)$ implies with a necessity that the condition \eqref{stabil_crit} does not hold for one of these roots, at least. For gauge theories, these considerations are valid for the physical modes only. The nonphysical modes can violate the conditions \eqref{stabil_crit}, \eqref{stabil_crit_gen}. Only when $\e_a(\omega)$ possesses singularities on the real axis in the $\omega$ plane may the additional stable branches in the dispersion law appear. However, these singularities cannot be obtained at any finite order in the derivative expansion.

In our case, we need to analyze the system of equations \eqref{eqm_lin_fl} for the longitudinal modes $\theta^\parallel_i$ at $b_3=0$. Assuming the zero boundary conditions at spatial infinity, the second equation in \eqref{eqm_lin_fl} gives
\begin{equation}
    \ddot{\theta}=-(\beta^4p_0'+b_4\Delta)^{-1}\big[\beta^4p_0'\dot{\theta}_0+(b_4-4b_7+2b_8)\dddot{\theta}_{\hspace{-0.11em}0}\big].
\end{equation}
Using this expression, we have for the first equation in \eqref{eqm_lin_fl},
\begin{multline}\label{zero_comp}
    2\be^{-2}\bigg\{\omega^4\Big[b_4+b_5+b_6+8b_7-4b_8-\frac{(b_4-4b_7+2b_8)^2}{\be^4p_0'-b_4\spk^2}\spk^2\Big]-\\
    -\omega^2\Big[c^{-2}\be^4p_0' +b_5\spk^2 +2\frac{\be^4p_0'\spk^2(b_4-4b_7+2b_8)}{\be^4p_0'-b_4\spk^2} \Big]+\frac{\be^8p_0'^2\spk^2}{\be^4p_0'-b_4\spk^2}\bigg\}\tilde{\theta}_0=0,
\end{multline}
where $\tilde{\theta}_0=\tilde{\theta}_0(\omega,\spk)$ is the Fourier transform of $\theta_0(\vk)$. Equation \eqref{zero_comp} has four roots in the complex $\omega$ plane. However, as we have already established, this is impossible for stable models. In order that equation \eqref{zero_comp} has only two nondegenerate roots, we have to demand
\begin{equation}\label{constr1}
    b_4+b_5+b_6+8b_7-4b_8=0,\qquad b_4-4b_7+2b_8=0.
\end{equation}
With the account of \eqref{constr1}, the condition \eqref{stabil_crit2} is satisfied as long as
\begin{equation}\label{b_5}
    b_5\leq0.
\end{equation}
The dispersion law takes the form
\begin{equation}\label{disp_law_fin}
    \omega^2=\frac{\be^8p_0'^2c^{2}_s\spk^2}{(b_4\spk^2-\be^4p_0')( -b_5c^{2}_s\spk^2-\be^4p_0')},
\end{equation}
whence it follows that
\begin{equation}\label{b_4}
    b_4\geq0.
\end{equation}
In the flat space-time limit, the dispersion law \eqref{disp_law_fin} reduces to $\omega^2=0$ as it should be. At the small momenta $\spk$, the terms with higher derivatives are negligible, and we have the standard linear dispersion law for phonons with the speed of sound $c_s$. At large momenta, $\omega(\spk)$ tends to zero provided that $b_4\neq0$ and $b_5\neq0$. Such a behaviour of the dispersion law is unphysical since, in that case, there are perturbations with a tiny energy (and a large momentum) possessing a negative inertial mass. We identify our fluid with the dark matter that does not display such a property. Therefore, either we set $b_4$ and/or $b_5$ to zero, or suppose that, at large momenta, where $\omega'(|\spk|)$ becomes negative, it is necessary to take into account the corrections to the pressure of the higher order in derivatives than in \eqref{p2}. These terms may correct the unphysical behaviour of the dispersion law at large momenta.

The general solution to the equations \eqref{b3} and \eqref{constr1} can be written as
\begin{equation}
    \left[
      \begin{array}{c}
        b_1 \\
        b_2 \\
        b_3 \\
        b_4 \\
        b_5 \\
        b_6 \\
        b_7 \\
        b_8 \\
      \end{array}
    \right]=\left[
              \begin{array}{ccccc}
                -1 & 0 & 0 & 0 & 0 \\
                2 & 1 & 0 & 0 & 0 \\
                0 & 0 & 0 & 0 & 0 \\
                0 & 0 & 0 & -2 & 0 \\
                0 & 0 & 1 & 0 & 0 \\
                0 & 0 & -1 & 6 & 0 \\
                0 & 0 & 0 & 0 & 1 \\
                0 & 0 & 0 & 1 & 2 \\
              \end{array}
            \right]
            \left[
              \begin{array}{c}
                c_1 \\
                c_2 \\
                c_3 \\
                c_4 \\
                c_5 \\
              \end{array}
            \right],
\end{equation}
where $c_i$ are arbitrary constants. Evidently, the inequalities \eqref{b_5} and \eqref{b_4} are equivalent to
\begin{equation}\label{c_ineq}
    c_3\leq0, \qquad c_4\leq0.
\end{equation}
As a result,
\begin{equation}
    p_2=2c_1G_{ab}t^at^b+c_2R_{ab}t^at^b+c_3\pr^{ab}a_aa_bt^2-2c_4(\nabla_au^a)^2t^2-c_5\nabla_a(\rho^{[a}\nabla_b\rho^{b]})t^4,
\end{equation}
where
\begin{equation}
    G_{ab}:=R_{ab}-\tfrac12h_{ab}R,\qquad\pr_{ab}:=h_{ab}-u_au_b.
\end{equation}
The case $b_4=0$ corresponds to $c_4=0$, and the case $b_5=0$ implies $c_3=0$. The inequalities \eqref{c_ineq} assure that the structures standing at these coefficients give nonnegative contributions to the pressure. Further constraints on the coefficients $c_i$ can be found, if one analyzes the stability of small perturbations on a curved background and nonuniform fluid flow. In this paper, we shall not conduct such a study. We should only note that the structures at $c_1$ and $c_2$ are of a fixed sign and, consequently, can be made nonnegative as long as the weak (for the structure at $c_1$) and strong (for the structure at $c_2$) energy conditions are fulfilled. The structure at $c_5$ has not a definite sign, and so one may surmise that a further investigation of stability will lead to $c_5=0$.

\section{Nonlinear corrections}\label{NonlinCorr}

In this section, we consider the quadratic in $\theta^a$ contributions to the equations of fluid motion in the limit \eqref{flat_lim}, when all the gradients of the metric can be neglected. Furthermore, we restrict our consideration to the leading in derivatives term \eqref{eqmot_0}. Primarily, we shall be interested in the influence of the nonlinear terms in the equations of motion on the dynamics of the transverse modes.

In order to find the quadratic contribution, we need to take a variation with respect to $x^\mu$ of the linearized equations of motion \eqref{E0_lin}. It is convenient to express this variation in terms of $\theta^a$ rather than $\de x^\mu$ just as it was done in deriving the linearized equations. The variations of all the structures entering the expression \eqref{E0_lin} are evaluated according to the rules \eqref{variations}, but we have to take into account that
\begin{equation}\label{scnd_var}
    \de\theta^a=-\theta^b\tilde{\nabla}_b\theta^a.
\end{equation}
Since the hydrodynamic equations in the Fock-Taub representation possess the property \eqref{vort_cons_law}, it is useful to develop in a Taylor series $\sqrt{|h|}E_c$ rather than $E_c$. Then
\begin{equation}
    \sqrt{|h|}E^0_c=\de(\sqrt{|h|}E^0_c)+\tfrac12\de^2(\sqrt{|h|}E^0_c)+\dots=\sqrt{|h|}(\de E^0_c+\de E^0_c\nabla_a\theta^a+\tfrac12\de^2E^0_c+\dots),
\end{equation}
where in the last equation we suppose, for simplicity, that the background fields satisfy the classical equations of motion \eqref{eqmot_0}. The ellipses mark the higher terms of the series in $\theta^a$. Performing rather bulky, but simple, calculations, in the limit \eqref{flat_lim}, we arrive at
\begin{multline}\label{eqm_sec_pert}
    \sqrt{|h|}E^0_c=-\frac{\partial}{\partial\tau}\bigg\{\big[2\be^2p_0'(1+\partial_a\theta^a)+4\be^4p_0''\dot{\theta}_0\big](\partial_c\theta_0+\dot{\theta}_c) +2\be^2p_0'\dot{\theta}^a\partial_c\theta_a+\\
    +\de^0_c\Big[\be^2p_0'\big(2\partial_a\theta^a+(\partial_a\theta^a)^2+\partial_b\theta^a\partial_a\theta^b\big) +2\be^4p_0''\big(2\dot{\theta}_0(1+\partial_a\theta^a)+\dot{\theta}_a\dot{\theta}^a\big) +4\be^6p_0'''\dot{\theta}_0^2 \Big] +\dots\bigg\}=0.
\end{multline}
Remark that $\rho^a$ described in \eqref{flat_lim} with $\be=const$ does provide the solution to the equations of motion not only in the limit when the gradients of the metric can be neglected, but also for an arbitrary stationary metric with the Killing vector $\rho^a$. Of course, the last equality in \eqref{flat_lim} does not hold for an arbitrary stationary metric. The field $\theta^a$ has the mass dimension $-1$. It follows from \eqref{eqm_sec_pert} that the amplitude of this field is small provided that
\begin{equation}\label{ampltd_sm}
    |\spk|\theta^a\sim\theta^a/\la_{ph}\ll1,
\end{equation}
where $\la_{ph}$ is a characteristic wavelength of phonons. To put it in another way, recollecting the definition of $\theta^a$, the deviation of fluid particles from their average positions should be small in comparison with the wavelength. Only in this case can the higher terms of the expansion \eqref{eqm_sec_pert} be thrown away.

Let us investigate the influence of the nonlinear terms on the transverse perturbations of the fluid. We shall solve equation \eqref{eqm_sec_pert} using the perturbation theory assuming that the amplitude of the perturbation $\theta^a$ is small. If the transverse perturbations are small then, as we saw, their dynamics are trivial on the background \eqref{flat_lim}. The small transverse modes can evolve in a nontrivial way on the background \eqref{flat_lim} only due to the presence of nonlinear contributions of the longitudinal modes with a sufficiently large amplitude. At that, the amplitude of the longitudinal modes has to satisfy \eqref{ampltd_sm} for we can use \eqref{eqm_sec_pert} for their description. In that case, the longitudinal wave of a sufficiently large amplitude induces small transverse perturbations. This mechanism for generation of the transverse perturbations is analogous to the acoustic Rayleigh streaming (see, e.g., \cite{LandLifshHyd,RudSol,Eckart,Wester,Lighthill}), but it does not need the viscosity of fluid to work. In fact, the vorticity \eqref{vorticity} is conserved in our case. For relativistic fluids, we have to distinguish the transverse and entropy-vortex perturbations in higher orders of perturbation theory (see equation \eqref{trans_pert_ind} below).

Let $A$ be an amplitude of the longitudinal modes $\bar{\theta}^\parallel_a$ and $a$ be an amplitude of the induced perturbations $\psi_a$ (both longitudinal and transverse). The total perturbation takes the form
\begin{equation}\label{pert_tot_quad}
    \theta_a=\bar{\theta}^\parallel_a+\psi_a=\bar{\theta}^\parallel_a+\psi^\parallel_a+\psi_a^\perp.
\end{equation}
Then we deduce from \eqref{eqm_sec_pert} that
\begin{equation}\label{ampltd_ind}
    |\spk|a\sim\spk^2A^2\ll|\spk|A.
\end{equation}
If, in the region of space at issue, the transverse perturbations are absent at the initial instant of time (that we shall assume) then their amplitude will be of the order \eqref{ampltd_ind} after the perturbation comes to the region considered. Therefore, the choice of $\bar{\theta}^\parallel_a$ obeying \eqref{prll_mod} as the zeroth order approximation to the solution of \eqref{eqm_sec_pert} is justified.

In the leading order, equation \eqref{eqm_sec_pert} is written as
\begin{equation}\label{eqm_nonlin}
    \de E^0_c\Big|_{\theta_a\rightarrow\psi_a}+\Big[\de E^0_c\nabla_a\theta^a+\tfrac12\de^2E^0_c\Big]_{\theta_a\rightarrow\bar{\theta}^\parallel_a}=:\de E^0_c\Big|_{\theta_a\rightarrow\psi_a}-\dot{f}_c=0,
\end{equation}
that is, the linear in $\theta_a$ terms in \eqref{eqm_sec_pert} (see also \eqref{eqm_lin_secord}) are replaced by $\psi_a$, and the quadratic terms are substituted for $\bar{\theta}^\parallel_a$. Employing \eqref{prll_mod}, we derive from \eqref{eqm_sec_pert},
\begin{equation}
\begin{split}
    f_0&=\be^2\Big\{p_0'\partial_{ij}\theta\partial_{ij}\theta-2\be^2p_0''\partial_i\dot{\theta}\partial_i\dot{\theta}+(\De\theta)^2\big[p_0'+c^2_s(4\be^2p_0'' +6p_0')+c^4_s(4\be^4p_0'''+14\be^2p_0''+8p_0') \big] \Big\},\\
    f_i&=2\be^2p_0'(c^2_s\De\theta\partial_i\dot{\theta}-\partial_j\dot{\theta}\partial_{ij}\theta),
\end{split}
\end{equation}
where $\bar{\theta}^\parallel_a=\partial_a\theta$. Splitting the linear part of \eqref{eqm_nonlin} as in \eqref{decomp_pot_vor}, we have for the transverse part of the perturbation at $c=i$,
\begin{equation}
    \partial_{[i}\ddot{\psi}^\perp_{j]}=\partial_\tau\big[\partial_{k[i}\dot{\theta}\partial_{j]k}\theta-c^2_s\De\partial_{[i}\theta\partial_{j]}\dot{\theta} \big].
\end{equation}
Taking into account that the fluid flow was unperturbed at the initial instant of time $\tau$, we come to
\begin{equation}\label{trans_pert_ind}
    \partial_{[i}\dot{\psi}^\perp_{j]}=\partial_{k[i}\dot{\theta}\partial_{j]k}\theta-c^2_s\De\partial_{[i}\theta\partial_{j]}\dot{\theta}.
\end{equation}
This equation describes the induced transverse perturbation moving along with the sound wave. Equation \eqref{trans_pert_ind} can be derived in a more straightforward manner from the vorticity potential conservation law \eqref{hydr_eqm}. In the case at issue, $r_a=t_a$ and
\begin{equation}\label{t_sord}
    t_a=\partial_a\theta_0+\dot{\theta}_a-2\dot{\theta}_0(\partial_a\theta_0+\dot{\theta}_a)+\dot{\theta}_c\partial_a\theta^c+\de^0_a(1-2\dot{\theta}_0-\dot{\theta}^2+4\dot{\theta}_0^2)+\dots,
\end{equation}
where the dots denote the terms of higher orders in $\theta^a$. Substituting \eqref{pert_tot_quad} into \eqref{t_sord}, taking the external derivative of $t_i$, and setting it to zero, we come to \eqref{trans_pert_ind}. Actually, equation \eqref{trans_pert_ind} is the zero vorticity condition accurate within the second order in $\theta^a$.

The potential part of the perturbation $\psi_a$ is specified by the system of equations
\begin{equation}
    2\be^2p_0'\big[c_s^{-2}\ddot{\psi}_0+\De\dot{\psi}\big]=\dot{f}_0,\qquad-2\be^2p_0'\De(\dot{\psi}_0+\ddot\psi)=\partial_i\dot{f}_i,
\end{equation}
where $\psi_i=\partial_i\psi$. If the fluid flow was unperturbed at the initial instant of time in the region of space considered then one derivative with respect to $\tau$ in the second equation can be ``canceled''. As a result, summing the equation obtained with the first one, we deduce
\begin{equation}\label{eqm_nonlin_pot}
    2\be^2p_0'\big[c_s^{-2}\ddot{\psi}_0-\De\psi_0\big]=\dot{f}_0+\partial_if_i,
\end{equation}
i.e., the wave equation with a source. This equation describes, in particular, the appearance of small overtones in a sound wave that have a doubled frequency as compared with the initial one. This phenomenon is well-known in acoustics (see, e.g., \cite{LandLifshHyd,RudSol}).

Equation \eqref{eqm_nonlin_pot} can be considerably simplified, if one keeps there only the leading term at $x\rightarrow0$ (see the asymptotics \eqref{asympt_wfl} and \eqref{sound_speed}). In this limit,
\begin{equation}
    c^{-2}_s\approx-2\be^2\frac{p_0''}{p_0'},
\end{equation}
and
\begin{equation}
\begin{aligned}
    \frac{f_0}{2\be^2p_0'}&\approx\frac12\Big[\Big(\frac{p_0'p_0'''}{(p_0'')^2}-1\Big)(\De\theta)^2-2\be^2\frac{p_0''}{p_0'}\partial_i\dot{\theta}\partial_i\dot{\theta}+\partial_{ij}\theta \partial_{ij}\theta \Big],&\quad
    \frac{f_i}{2\be^2p_0'}&\approx-\partial_j\dot{\theta}\partial_{ij}\theta,\\
    \frac{\dot{f}_0}{2\be^2p_0'}&\approx \Big(\frac{p_0'p_0'''}{(p_0'')^2}-1\Big)\De\theta\De\dot{\theta}+\partial_i\dot{\theta}\partial_i\De\theta+\partial_{ij}\theta\partial_{ij}\dot{\theta},&\quad \frac{\partial_i f_i}{2\be^2p_0'}&\approx-\partial_i(\partial_{ij}\theta\partial_j\dot{\theta}).
\end{aligned}
\end{equation}
Whence
\begin{equation}
    (\dot{f}_0+\partial_if_i)/(2\be^2p_0')\approx\De\theta\De\dot{\theta}\Big(\frac{p_0'p_0'''}{(p_0'')^2}-1\Big)=-\tfrac1{n}\De\theta\De\dot{\theta} =-\tfrac1{2n}\partial_\tau[(\De\theta)^2].
\end{equation}
Consequently,
\begin{equation}
    c_s^{-2}\ddot{\psi}_0-\De\psi_0=-\tfrac1{2n}\partial_\tau[(\De\theta)^2].
\end{equation}
These calculations show that the classical perturbation theory, i.e., in fact, the perturbation theory for averages, is applicable in the flat space-time limit $x\rightarrow0$ as long as the condition \eqref{ampltd_sm} holds. In Sect. \ref{QuanEff}, we shall see that the standard in-out quantum perturbation theory does not work in this limit.

Thus, the accounting of the nonlinear terms of second order in the fluid equations of motion does not lead to self-maintained nontrivial dynamics of the transverse perturbations. The nontrivial dynamics of the transverse modes may arise at the third order. Supposing that the connected Green function $\lan T\{\theta^a\theta^b\}\ran$ is not zero and neglecting the connected correlator of the third order in equations \eqref{eqmot_0} expanded up to the third order in $\theta_a$ and averaged over the vacuum state, one can derive a closed system of equations for the averaged field $\theta^a$ and the Green function $\lan T\{\theta^a\theta^b\}\ran$. This approach is equivalent to the leading approximation in the effective action formalism for composite operators \cite{CorJaTom}. In our case, the role of an independent composite operator is played by the two-point Green function. The study of this system of equations will be given elsewhere.

\section{Concrete form of the classical potential}\label{ConFormClPot}

In the previous sections, we have worked with a general expression \eqref{eff_pot} for the potential without any reference to its peculiar structure and the values of the coefficient entering it. In this section, we consider several particular cases of \eqref{eff_pot} and discuss the restrictions on the coefficients coming from the normalization conditions \eqref{cond_tem_zero}, \eqref{cond_ewmin}, and \eqref{cond_redshift}.

First of all observe that if we truncate the expansion \eqref{eff_pot} over $l$ at the terms with $l_{max}=L$ (recall we suppose that the anomalous scaling is small and so the coefficients at the logarithms are small as well) then the pressure will contain $3(L+1)$ structures $t^{2k}\ln^lt^2$. Hence, adjusting the coefficient in such a potential, we can make vanish no more than $(3L+2)$ derivatives with respect to $\rho^2$ at $\rho^2=1$. This is equivalent to a vanishing of $(3L+2)$ derivatives of the potential with respect to $t^2$ at $t^2=1$. As a result, in the weak field limit, the polytropic index is equal to [see \eqref{asympt_wfl}]
\begin{equation}
    n_{max}=3L+1.
\end{equation}
In all, the pressure \eqref{eff_pot} at $l_{max}=L$ contains $6(L+1)$ arbitrary constants. Therefore, usually, it is not a problem even for the maximal polytropic index $n_{max}$ to impose the additional normalization conditions \eqref{cond_ewmin} and \eqref{cond_redshift}. Nevertheless, having imposed the constraints \eqref{cond_ewmin} and \eqref{cond_redshift}, the natural electroweak scale arises in the potential -- the normalization conditions involve the vacuum expectation value of the Higgs field $\eta$ and the Higgs boson mass $m_\chi$. So, in order to satisfy the asymptotics \eqref{asympt_wfl} with the physical value \eqref{A_const} of the constant $A$ at small $n<8$, we have to fine-tune the constants $b_{kls}$ in such a way that the coefficients at $t^{2k}\ln^lt^2$ agree with the value \eqref{A_const} at $\phi=\eta$. We shall see this below for the concrete examples with $L=\{0,1,4\}$.

Let us start with the simplest model $L=0$. In this case, $n_{max}=1$ and coincides with the minimal value of $n$ admissible by the conditions 1 and 2 of Sect. \ref{QGA}. Imposing the normalization conditions \eqref{cond_tem_zero}, \eqref{cond_ewmin}, and \eqref{cond_redshift}, we arrive at
\begin{equation}
    p_0=-\frac{m^2_\chi}{8\eta^2}(|\phi|^2-\eta^2)^2-\frac{b_{100}}2(t^2-1)^2,
\end{equation}
i.e., the potential falls into two noninteracting parts -- the Higgs potential and the potential (the pressure) of the field $\xi^\mu$. The value of the constant $b_{100}$ is fixed by the condition \eqref{asympt_wfl},
\begin{equation}
    b_{100}=-A_1.
\end{equation}
The value of $A_1$ is given in \eqref{A_const_vls}.

Now we consider $L=1$. The model of this type with the additional quantum corrections was studied in \cite{gmse}. In this case, $n_{max}=4$ and we put $n=n_{max}$. Imposing the normalization conditions providing the polytropic index $n=4$ in \eqref{asympt_wfl} and the normalization conditions \eqref{cond_ewmin}, \eqref{cond_redshift} (eight conditions, in all), we obtain a four-parameter family of potentials
\begin{multline}
    p_0=-\frac{m^2_\chi}{8\eta^2}(|\phi|^2-\eta^2)^2+(|\phi|^2-\eta^2)\Big[\frac{b_{010}}{\eta^2} (t^2-1-\ln t^2)+\frac{b_{110}}{\eta^2} (t^2-1-t^2\ln t^2) -b_{012}\eta^2\big(t^2-1-\tfrac{|\phi|^2}{\eta^2}\ln t^2\big) \Big]-\\
    -b_{210}\Big[(t^2-1)\big(3t^2-2+5\tfrac{|\phi|^2}{\eta^2}\big)-(t^4+\tfrac{|\phi|^2}{\eta^2}(4t^2+1))\ln t^2 \Big].
\end{multline}
In particular, for $|\phi|=\eta$, we have
\begin{equation}
    p_0\big|_{|\phi|=\eta}=b_{210}[3-3t^4+(t^4+4t^2+1)\ln t^2]=\frac{b_{210}}{30}(t^2-1)^5+O\big((t^2-1)^6\big).
\end{equation}
The constant $b_{210}$ is fixed by the asymptotics \eqref{asympt_wfl} with the constant $A$ presented in \eqref{A_const_vls}.

As the last example, we consider the case $L=4$. In \cite{KalKaz1}, it was suggested that we can get rid of the dimensional constant $\mu^2$ in the Higgs potential (see the notation in \cite{gmse}) replacing this constant by $\al t^2$. Then, redefining $t\rightarrow\La_{dm}^{-1}t$, only two dimensional constants remain in the standard model with gravity: the gravitational constant and the cosmological one. Let us show that such a model (for definiteness, we call it conformal) can be realized at $L\geq4$. Namely, in the conformal model, the pressure \eqref{eff_pot} does not contain the dimensional constants, after the dilatation $t\rightarrow\La_{dm}^{-1}t$, and
\begin{equation}
    a_{kl}(|\phi|^2)=b_{kl}|\phi|^{4-2k}.
\end{equation}
In all, there are $3(L+1)$ arbitrary constants in the potential, the coefficient at $t^2|\phi|^2$ being of the order of the electroweak scale. Therefore, in order to arrive at the asymptotics \eqref{asympt_wfl}, \eqref{A_const}, we need to adjust the coefficients in the potential such that the polytropic index $n\geq8$. Taking into account the requirements \eqref{cond_ewmin}, \eqref{cond_redshift}, we have $(n+4)\geq12$ conditions. For $L=3$, the pressure contains 12 constants, but the system proves to be inconsistent. Actually, this system of conditions does not possess any solution already at $n\geq7$. Only at $L\geq4$ can we realize the conformal model.

Let us consider $L=4$. It turns out that the natural polytropic index is $n=8$ for this model. For $n\geq9$, having imposed the normalization conditions, the coefficients at the logarithmic corrections in \eqref{eff_pot} prove to be large in comparison with the same structures without logarithms. This contradicts to the requirement of a small anomalous scaling that we assume. For $n=8$, the normalization conditions \eqref{cond_ewmin}, \eqref{cond_redshift}, imply a three-parameter family of potentials
\begin{multline}
    p_0=-\frac{m^2_\chi\eta^2}{8}\Big[(r^2-t^2)^2 +\tfrac{1}{2}(7t^4-10t^2r^2+3r^4)\ln t^2 +\tfrac{1}{64}(79t^4+16t^2r^2-31r^4)\ln^2 t^2+\\
    +\tfrac{1}{96}(13t^2-88r^2)t^2\ln^3 t^2-\tfrac{r^4}{96}\ln^4t^2 \Big]  +b_{03}\eta^4\Big[24(r^2-t^2)^2\ln t^2+9(r^4-t^4)\ln^2 t^2 +(r^4-8t^2r^2+t^4)\ln^3 t^2\Big]+\\
    +b_{14}\eta^2\Big[48(r^2-t^2)^2\ln t^2+\tfrac38(35r^4+16t^2r^2-51t^4)\ln^2 t^2 -\tfrac94(8r^2-t^2)t^2\ln^3t^2-(r^4-t^2r^2)\ln^4t^2 \Big]+\\
    +b_{24} \Big[168(r^2-t^2)^2\ln t^2+(r^4-t^4)(48\ln^2 t^2-\ln^4t^2) -72r^2t^2\ln^3t^2 \Big],
\end{multline}
where $r:=|\phi|/\eta$. The asymptotics \eqref{asympt_wfl} with the constant $A$ given in \eqref{A_const_vls} fixes one of the three arbitrary constants. For instance,
\begin{equation}
    b_{03}=\frac{35}{57122}-\frac{7m^2}{384\eta^2}-\frac{3b_{14}}{\eta^2}-\frac{15b_{24}}{\eta^4},
\end{equation}
and then
\begin{multline}\label{P-0_conf}
    p_0=-\frac{m^2_\chi\eta^2}{8}(r^2-t^2)^2 +\eta^4\Big\{\big[\tfrac{420}{28561}(r^2-t^2)^2-\tfrac{m_\chi^2 r^2}{4\eta^2}(r^2-t^2)\big]\ln t^2 +\big[ (\tfrac{315}{57122}-\tfrac{53m_\chi^2}{512\eta^2})r^4 -\tfrac{m^2_\chi}{32}t^2r^2+\\
    +(\tfrac{5m_\chi^2}{512\eta^2}-\tfrac{315}{57122})t^4\big]\ln^2 t^2
    +\big[(\tfrac{35}{57122}-\tfrac{7m^2_\chi}{384\eta^2})r^4 +(\tfrac{m^2_\chi}{32\eta^2}-\tfrac{140}{28561})t^2r^2+(\tfrac{35}{57122}-\tfrac{m^2_\chi}{768\eta^2})t^4 \big]\ln^3 t^2-\tfrac{m_\chi^2r^4}{768\eta^2}\ln^4t^2 \Big\}-\\
    -b_{14}\eta^2\Big[24(r^2-t^2)^2\ln t^2+(\tfrac{111}{8}r^4-6t^2r^2-\tfrac{63}{8}t^4)\ln^2 t^2 +(3r^4-6t^2r^2+\tfrac34t^4)\ln^3t^2+(\tfrac{r^4}{4}-t^2r^2)\ln^4t^2 \Big]-\\
    -b_{24} \Big[192(r^2-t^2)^2\ln t^2+(r^4-t^4)(87\ln^2 t^2+\ln^4t^2) +(15r^4-48t^2r^2+15t^4)\ln^3t^2 \Big].
\end{multline}
One can check that the coefficients at the logarithms are small for reasonable, in particular, for vanishing, values of the constants $b_{12}$, $b_{24}$. Since the polytropic index is an even number, the domains with a negative energy density do not appear in the weak field limit.

The question as to make the gravitational sector of a theory invariant with respect to the global conformal (Weyl) transformations by substituting the Hilbert-Einstein action for the terms with second derivatives \eqref{p2} and the logarithmic corrections to them remains open. As we have already noted in Sect. \ref{QGA}, we may hope this theory is conformally invariant on the quantum level. This happens when the coefficients at the logarithms in \eqref{P-0_conf} coincide with the renormalization group functions $\be$ and $\ga$ taken on the cutoff scale $\La$. Of course, as the expression \eqref{P-0_conf} is approximate, the coincidence should be approximate too. The analysis above shows that the conformal model of a dark matter can be realized for the polytropic indices $n\geq8$ only. Therefore, it can be excluded by the experimental data regarding the properties of a dark matter.

\section{Some quantum effects}\label{QuanEff}

In this section, we try to estimate the decay rate of the Higgs boson to phonons and evaluate the one-loop contribution of phonons to the effective potential.

To begin with, we consider the one-loop correction to the effective action coming from the quantum fluctuations of phonons on the stationary background with the Killing vector $\xi^\mu$ in the leading order in derivatives. If the fluid rests in the adapted system of coordinates, where $\xi^\mu=(1,0,0,0)$, then the phonon dispersion law has the form
\begin{equation}
    g^{00}\omega^2+2c_sg^{0i}\omega k_i+c^2_sg^{ij}k_ik_j=0.
\end{equation}
In the flat space-time limit,
\begin{equation}
    \Ga^{(1)}=-\frac12\int dx^0\sum_\al^{\omega_\al<\La}\omega_\al=-\int d^4x\int\frac{d\spk}{2(2\pi)^3}\theta(\La-c_s|\spk|)c_s|\spk|=-\int d^4x c_s^{-3}\frac{\La^4}{16\pi^2}.
\end{equation}
When the derivatives of the metric can be neglected, the dependence on it is easily recovered (for details, see (18) in \cite{gmse})
\begin{equation}\label{1loop_phon}
    \Ga^{(1)}=-\int d^4x\sqrt{|g|} c_s^{-3}\frac{\La^4t^4}{16\pi^2}.
\end{equation}
The explicit expression for $c_s$ is given by formula \eqref{sound_speed}. The expression obtained is singular in the flat space-time limit when $c_s\rightarrow0$. In order to comply with the conditions 1 and 2 of Sect. \ref{QGA}, this contribution to the effective action must be completely canceled out. One can think that it is possible to cancel only the singular in the flat space-time limit part of the contribution \eqref{1loop_phon}. This can be done, if one adds to \eqref{1loop_phon} the terms proportional to
\begin{equation}
    (t^2-1)^{-3/2},\qquad (t^2-1)^{-1/2},\qquad (t^2-1)^{1/2},
\end{equation}
with the corresponding coefficients. However, upon stretching $t\rightarrow\La^{-1}t$ and developing as a series in $t^2$, these terms lead to the appearance of the coupling constants with arbitrary large negative mass dimension in the theory. This contradicts the property mentioned in Sect. \ref{QGA} after the condition 3 that we wish to preserve.

In conclusion, we consider the decay of the Higgs boson to phonons. As long as the pressure \eqref{eff_pot} contains the terms that depend on the product of the fields $\phi$ and $\xi^\mu$, one may suppose that this decay is possible. Then the experimental data concerning the Higgs boson decay can be used to impose the constraints on the values of the coefficients at the corresponding terms. However, we shall show now that, in the weak field limit, the standard in-out perturbation theory is not applicable to phonons since they constitute a condensate -- the macroscopic field $\xi^\mu$ -- and, in a certain sense, are not elementary fields of the theory in this limit.

Consider the case when the gradients of the metric can be neglected and the background field obeys \eqref{flat_lim}. We shall consider the longitudinal phonon modes only. It follows from \eqref{eqm_lin_secord} that the canonically normalized field is defined as
\begin{equation}
    \theta_c^0:=\Big(-\frac{2\be^2p_0'}{c_s^2}\Big)^{1/2}\theta^0=\Big(\frac{\e_0}{c_s^2}\Big)^{1/2}\theta^0,\qquad \theta_c^i:=\e_0^{1/2}\theta^i.
\end{equation}
The expressions for variations of the potential \eqref{eff_pot} with respect to $\theta^a$ are present in \eqref{variations} and \eqref{scnd_var}. It is only relevant for us that, upon variation, the field $\theta^a$ always enters with one derivative. Let us show that the Higgs boson decay to a larger number of phonons is more probable than to a lesser number of them. In other words, the decay process is not perturbative.

Indeed, according to the standard expression (see, e.g., \cite{BjoDre}) for the differential decay rate to $n$ identical particles in the rest frame of a decaying particle, we have
\begin{equation}\label{decay_prob}
    d\Ga(\chi\rightarrow n\theta)=(2\pi)^4\de^4\Big(p-\sum_{i=1}^nk_i\Big)\frac{|M|^2}{2m_\chi n!}\Big(\frac{c_s^2}{\e_0}\Big)^{n}\prod_{i=1}^n\frac{d\spk_i}{2\omega_i(2\pi)^3},
\end{equation}
where $p^\mu=(m_\chi,0,0,0)$ is the four-momentum of the Higgs boson, $\omega_i=c_s|\spk_i|$, and $M$ is the invariant scattering amplitude. Its square is of the order
\begin{equation}
    |M|^2\sim\eta^2\bigg[\Big(\frac{\partial}{\partial\rho^2}\Big)^n\frac{\partial p_0}{\partial|\phi|^2}\bigg]_{|\phi|=\eta,\rho^2=1}^2\prod_{i=1}^n(2\omega_i)^2,
\end{equation}
where $\omega_i$ come from the derivatives with respect to $\tau$ acting on the field $\theta^0$. To obtain the total decay rate for $\chi\rightarrow n\theta$, it is necessary to integrate \eqref{decay_prob} over all the phonon momenta. Roughly, this integral can be estimated as follows: every $\spk_i$ and $d\spk_i$ are replaced by $m_\chi/c_s$, and the delta-functions produce the factor $c_s^3m_\chi^{-4}$. Then
\begin{equation}
    \Ga(\chi\rightarrow n\theta)\sim \frac{\eta^2c_s^3}{m_\chi^5n!}\bigg[\Big(\frac{\partial}{\partial\rho^2}\Big)^n\frac{\partial p_0}{\partial|\phi|^2}\bigg]_{|\phi|=\eta,\rho^2=1}^2 \Big(\frac{m^4_\chi}{4\pi^3 c_s\e_0}\Big)^n.
\end{equation}
There also exits the contribution of the same order from the variation of $\sqrt{|h|}$ and expansion of the Higgs field $\chi(x(\vk))$ in $\theta^a$. Of course, the dependence on $n$ in this formula is incorrect since we did not evaluate the integrals exactly. Nevertheless, this formula implies that for
\begin{equation}\label{perturbativity}
    \frac{m^4_\chi}{4\pi^3 c_s\e_0}\gg1,
\end{equation}
the standard perturbation theory is not applicable. Taking into account the actual value of $\e_0$ in the weak field limit \eqref{dm_sun}, we see that to handle the decay process  $\chi\rightarrow n\theta$ by means of the standard $S$-matrix technique is a nonsense. The condition \eqref{perturbativity} agrees with the general strong coupling condition found in \cite{ENRW}.

In many respects, this situation resembles the theory of a quantum field with the field strength renormalization constant $Z\rightarrow0$ (see, for details, \cite{WeinbergB}). The kinetic term disappears from the action at $Z\rightarrow0$ for that field, just as in our model of phonons. In this sense the phonons are not elementary fields of the theory in the weak field limit. Only in the strong gravitational fields do the quantum properties of these particles become relevant.

On the other hand, we saw in Sect. \ref{NonlinCorr} that the classical perturbation theory, i.e., the leading order of the perturbation theory for averages, works even in the weak field limit provided \eqref{ampltd_sm} is fulfilled. This allows us to hope that the properly formulated in-in perturbation theory can be applied to describe quantum dynamics of the field $\xi^\mu$. In particular, it follows from \eqref{eqm_lin_secord} and \eqref{prll_mod} that, in the leading order, the wave packet of Higgs bosons influences the field $\theta^a$ solely though the sound speed $c_s$ due to small variations of the vacuum expectation value $\eta$. So, we have for the perturbation of the phonon potential $\de\theta$ over the background $\theta$,
\begin{equation}
    c_s^{-2}\dddot{\de\theta}-\Delta\de\dot{\theta}=-(\de c_s^{-2})\dddot{\theta}.
\end{equation}
Consequently, if the phonons are absent at the initial instant of time, they will not be produced later on. In fact, this means that the Higgs boson and the phonons are not mixed.

\section{Conclusion}

So, we have constructed a self-consistent formalism of quantum field theory on a curved background that solves dynamically the problem of dependence of observables on a choice of the vacuum state for quantum fields. At the same time, we have to state that many hopes formulated in \cite{gmse} regarding the predictive power of the theory developed are not justified because of the perturbative nonrenormalizability of relativistic hydrodynamics and its non-perturbativity at large momenta. A virtual lack of reliable experimental data that could reveal the additional peculiarities (symmetries) of the fluid effective action leaves a large freedom in a selection of the concrete model. However, we saw that, in the weak gravitational field limit, one can establish certain properties of this model that allow one to falsify it so long as this fluid is identified with a considerable part of a dark matter. Moreover, we can use the derivative expansion and construct the standard perturbation theory for processes with the energies satisfying the inequality inverse to \eqref{perturbativity}, where $m_\chi$ ought to be replaced by the energy of a process. In particular, it would be interesting to consider the decay of a low energy graviton to phonons when this process is perturbative. This process describes the influence of the dark matter on the graviton dynamics and seems result in a finite lifetime of the graviton. The classical perturbation theory is applicable under the much weaker condition \eqref{ampltd_sm}.

In order to identify the quantum gravitational anomaly with a dark matter or its considerable part, further investigations regarding the refinement and interpretation of the astrophysical data are also needed. This will allows us to find the universal constants $A$ and $n$ characterizing the fluid equation of motion in the limit of a weak gravitational field, or to prove that the dark matter is not described in this limit by the universal polytropic equation of state. The latter will signify that the hypothesis about the identification of the quantum gravitational anomaly with the dark matter is not valid. This, however, will not mean that the gravitational anomaly is absent. The relativistic fluid described by the field $\xi^\mu$ arises with a necessity in quantum field theory with gravity provided that the basic principles of quantum field theory and general relativity hold and the general covariance (the background independence) remains intact on a quantum level, i.e., the general covariance is a fundamental symmetry of Nature. As we have already mentioned, this is the most natural solution to the problem of time in quantum gravity or, put another way, the problem of a unique representation of the algebra of observables (other similar models see in \cite{Isham,IshKuch,KuchTor,KuchBrow}). If something can be called the flow of a physical (not coordinate) time then the vector field $\xi^\mu$ would be the most natural candidate to this role.

\paragraph{Acknowledgments.}

The work is supported in part by the Tomsk State University Academic D.~I. Mendeleev Fund Program and by the RFBR grant No. 13-02-00551.

\appendix
\section{Linearized equations of motion}\label{LinEQM}

In this appendix, we linearize the expressions \eqref{eqmot_18} with respect to the small perturbation $\theta^a$. In fact, as we have already mentioned in the main text, we need to evaluate the Lie derivative of the expressions \eqref{eqmot_18} using formulas \eqref{variations}. The outcome of these calculations is as follows
\[
\begin{split}
  \de E^1_c&=2\nabla_a\big[t^at^bR(\nabla_{(c}\theta_{b)}-4t_c\rho^d\nabla_b\theta_d)+t^at_c\theta^d\nabla_dR\big]+2t_ct^aR\nabla_{ab}\theta^b-\theta^a\nabla_aR\nabla_ct^2+2R\nabla_c(t^at^b)\nabla_a\theta_b,\\
  \de E^2_c&=4\nabla_d\big\{t^at^bt^d\big[(\theta^s\nabla_sR_{ab}+2R_{sb}\nabla_a\theta^s)\rho_c+R_{ab}\rho^s(\nabla_{(c}\theta_{s)}-6\rho_cg^d\nabla_d\theta_s ) \big]\big\} -2\nabla_b\big[t^at^b (\theta^s\nabla_sR_{ac}+\\
  &+R_{sc}\nabla_a\theta^s+R_{as}\nabla_c\theta^s-4R_{ab}t^d\rho^s\nabla_d\theta_s)\big] -(\theta^s\nabla_sR_{ab}+2R_{sb}\nabla_a\theta^s-4R_{ab}t^d\rho^s\nabla_d\theta_s) \nabla_c(t^at^b)-\\
  &-2R_{ac}t^at^b\nabla_{bs}\theta^s +4R_{ab}t^at^b\big(\nabla_c(t^d\rho^s)\nabla_d\theta_s +t^d\rho_c\nabla_{ds}\theta^s\big),\\
  \de E^3_c&=2\bigg\{\nabla_ct^b\Big[\nabla^2t^a\nabla_{(a}\theta_{b)}-2\nabla^2(t_bt^a\rho^s\nabla_a\theta_s)-\nabla^a(\nabla_{(a}\theta_{s)}\nabla^st_b)+\nabla^at^b\nabla_{as}\theta^s+\\ &+\nabla^at^s(\theta^dR_{d(sba)}+\nabla_{(sa)}\theta_b)+\nabla^s(\nabla_{sd}\theta_b+\theta^aR_{asbd})g^d\Big] -2\nabla^2t_b\nabla_c(t^bt^s\rho^d)\nabla_s\theta_d-\\
  &-2t^a\nabla_{as}\theta^s t^b\nabla^2t_b\rho_c +t^a\nabla_{as}\theta^s\nabla^2t_c+\nabla_a\bigg[t^a\Big(\nabla_{(c}\theta_{b)}\nabla^2t^b-2t^b\rho^s\nabla_b\theta_s\nabla^2t_c-2t^b\nabla^2t_b\nabla_{(c}\theta_{s)}\rho^s-\\ &-\nabla^s(\nabla_{(s}\theta_{b)}\nabla^bt_c) -2\nabla^2(t_ct^s\rho^b\nabla_s\theta_b)+\nabla_{bs}\theta^s\nabla^bt_c+\nabla^st^d(\theta^bR_{b(dcs)}+\nabla_{(ds)}\theta_c)+\\ &+\nabla^b(\nabla_{bs}\theta_c+\theta^dR_{dbcs})t^s \Big) -2t^at^b\rho_c\Big[\nabla_{sd}\theta^d\nabla^st_b-\nabla^s(\nabla_{(s}\theta_{d)}\nabla^dt_b) -2\nabla^2(t_bt^s\rho^d\nabla_s\theta_d)-\\
  &-4t^d\rho^s\nabla_d\theta_s\nabla^2t_b +\nabla_{(b}\theta_{s)}\nabla^2t^s +\nabla^st^d(\theta^rR_{r(dbs)}+\nabla_{(ds)}\theta_b)+\nabla^s(\nabla_{sd}\theta_b+\theta^rR_{rsbd})t^d \Big]\bigg] \bigg\},\\
  \de E^4_c&=4\Big\{\nabla_{ca}\big[\tfrac12t^at^b\nabla_{bs}\theta^s -t^a\nabla_b(t^bt^s\rho^d\nabla_s\theta_d)\big] -\nabla_c(t^at^s\rho^d\nabla_{ab}t^b)\nabla_s\theta_d -\nabla_s\Big[t^st^a\big(\rho^d\nabla_{ab}t^b\nabla_{(c}\theta_{d)}-\\
  &-4\rho_c\nabla_{ab}t^bt^d\rho^r\nabla_d\theta_r -2\rho_c\nabla_{ab}(t^bt^d\rho^r\nabla_d\theta_r) +\rho_c\nabla_a(t^b\nabla_{bd}\theta^d)\big)\Big] -t^at^d\rho_c\nabla_{ab}t^b\nabla_{ds}\theta^s \Big\},\\
\end{split}
\]\\[-2em]
\begin{equation}\label{lin_eqm_ap}
\begin{split}
  \de E^5_c&=-2\Big\{\nabla_ct\big[\nabla^2(t^au^b\nabla_a\theta_b)+\nabla^b(\nabla_{(a}\theta_{b)}\nabla^at)-\nabla^at\nabla_{ab}\theta^b \big]  +\nabla^2t\big[\nabla_c(t^au^b)\nabla_a\theta_b +t^au_c\nabla_{ab}\theta^b \big]+\\ &+\nabla_a\big[\nabla^2t(t^au^b\nabla_{(c}\theta_{b)}-3t^au_ct^d\rho^s\nabla_d\theta_s)-t^au_c\big(\nabla^2(t^d\rho^s\nabla_d\theta_s)+\nabla^d(\nabla_{(s}t_{d)}\nabla^st) -\nabla^dt\nabla_{ds}\theta^s \big) \big] \Big\},\\
  \de E^6_c&=-2\Big\{\nabla_c\big[\omega_at^at^b\nabla_b(t^d\rho^s\nabla_d\theta_s)-2(\omega_at^a)^2t^d\rho^s\nabla_d\theta_s\big] -(t^d\rho_c\nabla_{ds}\theta^s-t^d\rho^s\nabla_{cd}\theta_s) \big[\nabla_b(t^b\omega_at^a)+2(\omega_at^a)^2\big]-\\ &-\nabla_d\Big[t^d\rho^s(\nabla_{(c}\theta_{s)}-6t^r\rho_c\nabla_s\theta_r)\big(\nabla_b(t^b\omega_at^a)+2(\omega_at^a)^2\big)+t^d\rho_ct^b\nabla_{bs}\theta^s\omega_at^a +\nabla_b\big[t^bt^a\nabla_a(t^d\rho^s\nabla_d\theta_s)\big] \Big] \Big\},\\
  \de E^7_c&=\nabla_c\Big\{4\rho^a\rho^b\big[\nabla_{ab}(t^2t^dt^s\nabla_d\theta_s)+\nabla^dt^4(\nabla_{ab}\theta_d+\theta^sR_{sbda})\big]\Big\}+4\nabla_{ab}\big[ \rho^a\rho^b\nabla_c(t^2t^dt^s\nabla_d\theta_s)\big]-\\
  &-\nabla_{as}\theta^s\nabla_b(\rho^a\rho^b\nabla_ct^4)+\nabla_b(\rho^a\rho^b\nabla_dt^4)(\nabla_{ca}\theta^d+R^d_{\ asc}\theta^s) -\nabla_a\big[\rho^a\rho^b\nabla_ct^4\nabla_{bs}\theta^s+\\
  &+\rho^d\rho^b\nabla_ct^4(\nabla_{bd}\theta^a-R^a_{\ bds}\theta^s)-\rho^a\rho^b\nabla_dt^4(\nabla_{cb}\theta^d+R^d_{\ bsc}\theta^s)\big] +4\rho^a\rho^b\nabla_r(t^2t^dt^s\nabla_d\theta_s)R^r_{\ acb}-\\
  &-\rho^a\rho^b\nabla_dt^4\mathcal{L}_\theta R^d_{\ acb} +4\nabla_d\Big\{t^2 t^d t^s\nabla_{(c}\theta_{s)}\nabla_{ab}(\rho^a\rho^b) -6t^dt_ct^rt^s\nabla_d\theta_s\nabla_{ab}(\rho^a\rho^b) +t^2t^dt_c\big[\nabla_{as}\theta^s\nabla_b(\rho^a\rho^b)+\\
  &+\nabla_a(\rho^a\rho^b\nabla_{bs}\theta^s+\rho^b\rho^s(\nabla_{bs}\theta^a-R^a_{\ bsr}\theta^r)) \big] \Big\} +4t^2t^dt_c\nabla_{ds}\theta^s\nabla_{ab}(\rho^a\rho^b)-4t^2t^dt^s\nabla_{cd}\theta_s\nabla_{ab}(\rho^a\rho^b),\\
  \de E^8_c&=-4\Big\{\nabla_c\big[\rho^b\nabla_b(\rho^a\nabla_a(t^2t^dt^s\nabla_d\theta_s))\big]+\nabla_d\big[t^2t^dt^s\nabla_a(\rho^a\nabla_b\rho^b)(\nabla_{(c}\theta_{s)}  -6t^r\rho_c\nabla_r\theta_s)+\\
  &+t^2t^dt_c\nabla_a(\rho^a\rho^b\nabla_{bs}\theta^s) +t^dt_ct^a\nabla_{as}\theta^s\nabla_b\rho^b \big] +t^2t^d(t_c\nabla_{ds}\theta^s-t^s\nabla_{cd}\theta_s)\nabla_a(\rho^a\nabla_b\rho^b)\Big\}.
\end{split}
\end{equation}
All the calculations were carried out by hand and so the mistakes are possible, but unlikely.  As expected, the first two expressions vanish at $R=0$ and $R_{ab}=0$, respectively.


\begin{thebibliography}{999}

\bibitem{GorbRub1}
D.~S. Grobunov, V.~A. Rubakov,
\textsl{Introduction to the Theory of the Early Universe: Hot Big Bang Theory}
(World Scientific, London, 2011).

\bibitem{Schneidr}
P. Schneider,
\textsl{Extragalactic Astronomy and Cosmology}
(Springer, Heidelberg, 2015).

\bibitem{Odints}
S. Nojiri, S.~D. Odintsov,
Unified cosmic history in modified gravity: from $F(R)$ theory to Lorentz non-invariant models,
Phys. Rep. \textbf{505}, 59 (2011).

\bibitem{Milgr}
M. Milgrom,
The MOND paradigm,
arXiv:0801.3133.

\bibitem{Sanders}
R.~H. Sanders,
\textsl{The Dark Matter Problem: A Historical Perspective}
(Cambridge University Press, Cambridge, 2010).

\bibitem{FamaMcGau}
B. Famaey, S.~S. McGaugh,
Modified Newtonian dynamics (MOND): Observational phenomenology and relativistic extensions,
Living Rev. Relativity \textbf{15}, 10 (2012).

\bibitem{olep}
P.~O. Kazinski,
One-loop effective potential of the Higgs field on the Schwarzschild background,
Phys. Rev. D \textbf{80}, 124020 (2009).

\bibitem{KazShip}
P.~O. Kazinski, M.~A. Shipulya,
One-loop omega-potential of quantum fields with ellipsoid constant-energy surface dispersion law,
Ann. Phys. (NY) \textbf{326}, 2658 (2011).

\bibitem{gmse}
P.~O. Kazinski,
Gravitational mass-shift effect in the standard model,
Phys. Rev. D \textbf{85}, 044008 (2012).

\bibitem{prop}
P.~O. Kazinski,
Propagator of a scalar field on a stationary slowly varying gravitational background,
arXiv:1211.3448.

\bibitem{KalKaz1}
I.~S. Kalinichenko, P.~O. Kazinski,
High-temperature expansion of the one-loop free energy of a scalar field on a curved background,
Phys. Rev. D \textbf{87}, 084036 (2013).

\bibitem{KalKaz2}
I.~S. Kalinichenko, P.~O. Kazinski,
Non-perturbative corrections to the one-loop free energy induced by a massive scalar field on a stationary slowly varying in space gravitational background,
JHEP \textbf{1408}, 111 (2014).

\bibitem{AlvaWitt}
L. Alvarez-Gaum\'{e}, E. Witten,
Gravitational anomalies,
Nucl. Phys. B \textbf{234}, 269 (1984).

\bibitem{DeWGAQFT}
B.~S. DeWitt,
\textsl{The Global Approach to Quantum Field Theory} Vol. 1,2
(Clarendon Press, Oxford, 2003).

\bibitem{BuchOdinShap}
I.~L. Buchbinder, S.~D. Odintsov, and I.~L. Shapiro,
\textsl{Effective Action in Quantum Gravity}
(IOP, Bristol, 1992).

\bibitem{DeWQG1}
B.~S. DeWitt,
The quantization of geometry,
in L. Witten, ed., Gravitation: An Introduction to Current Research, Wiley, New York (1962).

\bibitem{DeWQG2}
B.~S. DeWitt,
Quantum theory of gravity. I. The canonical theory,
Phys. Rev. \textbf{160}, 1113 (1967).

\bibitem{MamMostStar}
S.~G. Mamaev, V.~M. Mostepanenko, and A.~A. Starobinskii,
Particle creation from the vacuum near a homogeneous isotropic singularity,
Zh. Eksp. Teor. Fiz. \textbf{70}, 1577 (1976)
[Sov. Phys. JETP \textbf{43}, 823 (1976)].

\bibitem{GriMaMos}
A.~A. Grib, S.~G. Mamayev, and V.~M. Mostepanenko,
\textsl{Vacuum Quantum Effects in Strong Fields}
(Friedmann Lab. Publ., St. Petersburg, 1994).

\bibitem{IshKuch}
C.~J. Isham, K.~V. Kucha\v{r},
Representations of spacetime diffeomorphisms. I and II,
Ann. Phys. (NY) \textbf{164}, 288 (1985); \textbf{164}, 316 (1985).

\bibitem{KuchTor}
K.~V. Kucha\v{r}, C.~G. Torre,
Gaussian reference fluid and interpretation of quantum geometrodynamics,
Phys. Rev. D \textbf{43}, 419 (1991).

\bibitem{Isham}
C.~J. Isham,
Canonical quantum gravity and the problem of time,
arXiv:gr-qc/9210011.

\bibitem{ConnRov}
A. Connes, C. Rovelli,
Von Neumann algebra automorphisms and time-thermodynamics relation in general covariant quantum theories,
Class. Quantum Grav. \textbf{11}, 2899 (1994).

\bibitem{KuchBrow}
J.~D. Brown, K.~V. Kucha\v{r},
Dust as a standard of space and time in canonical quantum gravity,
Phys. Rev. D \textbf{51}, 5600 (1995).

\bibitem{DeWQFTcspt}
B.~S. DeWitt,
Quantum field theory in curved spacetime,
Phys. Rep. \textbf{19}, 295 (1975).

\bibitem{BirDav}
N.~D. Birrel, P.~C.~W. Davies,
\textsl{Quantum Fields in Curved Space}
(Cambridge University Press, Cambridge, 1982).

\bibitem{FrolNov}
V.~P. Frolov, I.~D. Novikov,
\textsl{Black Hole Physics: Basic Concepts and New Developments}
(Springer-Verlag, New York, 1998).

\bibitem{Boulware}
D.~G. Boulware,
Quantum field theory in Schwarzschild and Rindler spaces,
Phys. Rev. D \textbf{11}, 1404 (1975).

\bibitem{HartHawk}
J.~B. Hartle, S.~W. Hawking,
Path-integral derivation of black-hole radiance,
Phys. Rev. D \textbf{13}, 2188 (1976).

\bibitem{Israel}
W. Israel,
Thermo-field dynamics of black holes,
Phys. Lett. A \textbf{57}, 107 (1976).

\bibitem{Unruh}
W.~G. Unruh,
Notes on black-hole evaporation,
Phys. Rev. D \textbf{14}, 870 (1976).

\bibitem{Haagb}
R. Haag,
\textsl{Local Quantum Physics: Fields, Particles, Algebras}
(Springer, Berlin, 1996).

\bibitem{DiEfGaNe}
M. Dineykhan, G.~V. Efimov, G. Ganbold, and S.~N. Nedelko,
\textsl{Oscillator Representation in Quantum Physics}
(Springer, Berlin, 1995).

\bibitem{Collins}
J.~C. Collins,
\textsl{Renormalization}
(Cambridge University Press, Cambridge, 1984).

\bibitem{CollinsPhys}
J. Collins, A. Perez, D. Sudarsky, L. Urrutia, and H. Vucetich,
Lorentz invariance and Quantum Gravity: An additional fine-tuning problem?,
Phys. Rev. Lett. \textbf{93}, 191301 (2004).

\bibitem{ColPerSud}
J. Collins, A. Perez, and D. Sudarsky,
Lorentz invariance violation and its role in quantum gravity phenomenology,
arXiv:hep-th/0603002.

\bibitem{Emch}
G.~G. Emch,
\textsl{Algebraic Methods in Statistical Mechanics and Quantum Field Theory}
(Wiley-Interscience, New York, 1972).

\bibitem{BogolShir}
N.~N. Bogolyubov, D.~V. Shirkov,
\textsl{Introduction to the Theory of Quantized Fields}
(Wiley, New York, 1980).

\bibitem{ZelnProc}
A.~I. Zel'nikov,
The vacuum polarization of massive fields in algebraically special spaces,
Abstracts of VI-th Soviet Gravitational Conference, Moscow, 1984,
edited by V.~N. Ponomareva
(Moscow State Pedagogical Insitute, Moscow, 1984),
p. 197 [in Russian].

\bibitem{GavrGit}
S.~P. Gavrilov, D.~M. Gitman,
Vacuum instability in external fields,
Phys. Rev. D \textbf{53}, 7162 (1996).

\bibitem{Page}
D.~N. Page,
Thermal stress tensors in static Einstein spaces,
Phys. Rev. D \textbf{25}, 1499 (1982).

\bibitem{Howard}
K.~W. Howard,
Vacuum $\lan T_\mu^{\ \nu}\ran$ in Schwarzschild spacetime,
Phys. Rev. D \textbf{30}, 2532 (1984).

\bibitem{BrOtPa}
M.~R. Brown, A.~C. Ottewill, and D.~N. Page,
Conformally invariant quantum field theory in static Einstein space-times,
Phys. Rev. D \textbf{33}, 2840 (1986).

\bibitem{FrZel}
V.~P. Frolov, A.~I. Zel'nikov,
Killing approximation for vacuum and thermal stress-energy tensor in static space-times,
Phys. Rev. D \textbf{35}, 3031 (1987).

\bibitem{AndHisSam}
P.~R. Anderson, W.~A. Hiscock, and D.~A. Samuel,
Stress-energy tensor of quantized scalar fields in static spherically symmetric spacetimes,
Phys. Rev. D \textbf{51}, 4337 (1995).

\bibitem{Weinb}
S. Weinberg,
Phenomenological Lagrangians,
Physica A \textbf{96}, 327 (1979).

\bibitem{DonGolHol}
J.~F. Donoghue, E. Golowich, and B.~R. Holstein,
\textsl{Dynamics of the Standard Model}
(Cambridge University Press, Cambridge, 1994).

\bibitem{ENRW}
S. Endlich, A. Nicolis, R. Rattazzi, and J. Wang,
The quantum mechanics of perfect fluids,
JHEP \textbf{1104}, 102 (2011).

\bibitem{Ventwoloop}
A.~E.~M. van de Ven,
Two-loop quantum gravity,
Nucl. Phys. B \textbf{378}, 309 (1992).

\bibitem{Torri}
G. Torrieri,
Viscosity of an ideal relativistic quantum fluid: A perturbative study,
Phys. Rev. D \textbf{85}, 065006 (2012).

\bibitem{DHNS}
S. Dubovsky, L. Hui, A. Nicolis, and D.~T. Son,
Effective field theory for hydrodynamics: thermodynamics, and the derivative expansion,
Phys. Rev. D \textbf{85}, 085029 (2012).

\bibitem{BallBell}
G. Ballesteros, B. Bellazzini,
Effective perfect fluids in cosmology,
JCAP \textbf{1304}, 001 (2013).

\bibitem{GripSuth}
B. Gripaios, D. Sutherland,
The quantum theory of fluids,
arXiv:1406.4422.

\bibitem{Ballest}
G. Ballesteros,
The effective theory of fluids at NLO and implications for dark energy,
arXiv:1410.2793.

\bibitem{KLSh}
P.~O. Kazinski, S.~L. Lyakhovich, and A.~A. Sharapov,
Lagrange structure and quantization,
JHEP \textbf{0507}, 076 (2005).

\bibitem{Taub}
A.~H. Taub,
General relativistic variational principle for perfect fluids,
Phys. Rev. \textbf{94}, 1468 (1954).

\bibitem{FockB}
V.~A. Fock,
\textsl{The Theory of Space, Time and Gravitation}
(Pergamon Press, London, 1959).

\bibitem{Schutz}
B.~F. Schutz, Jr.,
Perfect fluids in general relativity: velocity potentials and a variational principle,
Phys. Rev. D \textbf{2}, 2762 (1970).

\bibitem{Ray}
J.~R. Ray,
Lagrangian density for perfect fluids in General Relativity,
J. Math. Phys. \textbf{13}, 1451 (1972).

\bibitem{KiSmGo}
J. Kijowski, A. Sm\'{o}lski, and A. G\'{o}rnika,
Hamiltonian theory of self-gravitating perfect fluid and a method of effective deparametrization of Einstein's theory of gravitation,
Phys. Rev. D \textbf{41}, 1875 (1990).

\bibitem{Brown1}
J.~D. Brown,
Action functionals for relativistic perfect fluids,
Class. Quantum Grav. \textbf{10}, 1579 (1993).

\bibitem{Brown2}
J.~D. Brown,
On variational principles for gravitating perfect fluids,
arXiv:gr-qc/9407008.

\bibitem{HajKij}
P. H\'{a}j\'{i}\v{c}ek, J. Kijowski,
Lagrangian and Hamiltonian formalism for discontinuous fluid and gravitational field,
Phys. Rev. D \textbf{57}, 914 (1998);
\textbf{61} 129901(E) (2000) (erratum).

\bibitem{rrmm}
P.~O. Kazinski,
Radition reaction of multipole moments,
Zh. Eksp. Teor. Fiz. \textbf{132}, 370 (2007)
[J. Exp. Theor. Phys. \textbf{105}, 327 (2007)].

\bibitem{Mitskev}
N.~V. Mitskevich, A.~P. Efremov, and A.~I. Nesterov,
\textsl{Dynamics of Fields in General Relativity}
(Energoatomizdat, Moscow, 1985) [in Russian].

\bibitem{LandLifshCTF}
L.~D. Landau, E.~M. Lifshitz,
\textsl{The Classical Theory of Fields}
(Butterworth-Heinemann, San Francisco, 1994).

\bibitem{Zelman}
A.~L. Zelmanov,
Chronometric invariants and comoving coordinates in General Relativity,
Dokl. Akad. Nauk SSSR \textbf{107}, 815 (1956).

\bibitem{Moller}
C. M{\o}ller,
\textsl{The Theory of Relativity}
(Clarendon, Oxford, 1972).

\bibitem{Vladimir}
Yu.~S. Vladimirov,
\textsl{Reference Frames in Theory of Gravity}
(Energoizdat, Moscow, 1982) [in Russian].

\bibitem{HeTe}
M. Henneaux, C. Teitelboim,
\textsl{Quantization of Gauge Systems}
(Princeton University Press, Princeton, New Jersey, 1992).

\bibitem{LandLifshHyd}
L.~D. Landau, E.~M. Lifshitz,
\textsl{Fluid Mechanics}
(Pergamon, Oxford, 1987).

\bibitem{LandLifstat}
E.~M. Lifshits, L.~P. Pitaevskii,
\textsl{Statistical Physics. Part II}
(Pergamon, New York, 1980).

\bibitem{ColemB}
S. Coleman,
\textsl{Aspects of Symmetry}
(Cambridge University Press, Cambridge, 1985).

\bibitem{Hored}
G.~P. Horedt,
\textsl{Polytropes: Applications in Astrophysics and Related Fields}
(Kluwer, New York, 2004).

\bibitem{BAMoNo}
A. Balaguera-Antol\'{\i}nez, D.~F. Mota, and M. Nowakowski,
Astrophysical configurations with background cosmology: probing dark energy at astrophysical scales,
Mon. Not. R. Astron. Soc. \textbf{382}, 621 (2007).

\bibitem{ChernUFN}
A.~D. Chernin,
Dark energy and universal antigravitation,
Phys. Usp. \textbf{51}, 253 (2008).

\bibitem{MerBisTar}
M. Merafina, G.~S. Bisnovatyi-Kogan, and S.~O. Tarasov,
A brief analysis of self-gravitating polytropic models with a non-zero cosmological constant,
Astron. Astrophys. \textbf{541}, A84 (2012).

\bibitem{GarLiReLa}
S. Garbari, C. Liu, J.~I. Read, and G. Lake,
A new determination of the local dark matter density from the kinematics of K dwarfs,
Mon. Not. R. Astron. Soc. \textbf{425}, 1445 (2012).

\bibitem{ZRVLZ}
L. Zhang, H.-W. Rix, G. van de Ven, J. Bovy, C. Liu, and G. Zhao,
The gravitational potential near the Sun from SEGUE K-dwarf kinematics,
Astrophys. J. \textbf{772}, 108 (2013).

\bibitem{Read}
J.~I. Read,
The local dark matter density,
J. Phys. G \textbf{41}, 063101 (2014).

\bibitem{KSLBN}
P.~R. Kafle, S. Sharma, G.~F. Lewis, and J. Bland-Hawthorn,
On the shoulders of giants: Properties of the stellar halo and the Milky Way mass distribution,
Astrophys. J. \textbf{794}, 59 (2014).

\bibitem{BCMS}
C.~M. Bidin, G. Carraro, R.~A. M\'{e}ndez, and R. Smith,
Kinematical and chemical vertical structure of the galactic thick disk. II. A lack of dark matter in the solar neighborhood,
Astrophys. J. \textbf{751}, 30 (2012).

\bibitem{NaFrWhit}
J.~F. Navarro, C.~S. Frenk, and S.~D.~M. White,
The structure of cold dark matter halos,
Astrophys. J. \textbf{462}, 563 (1996).

\bibitem{DGSFMWGGKW}
F. Donato, G. Gentile, P. Salucci, C.~F. Martins, M.~I. Wilkinson, G. Gilmore, E.~K. Grebel, A. Koch, and R. Wyse,
A constant dark matter halo surface density in galaxies,
Mon. Not. R. Astron. Soc. \textbf{397}, 1169 (2009).

\bibitem{SaxFer}
C.~J. Saxton, I. Ferreras,
Polytropic dark haloes of elliptical galaxies,
Mon. Not. R. Astron. Soc. \textbf{405}, 77 (2010).

\bibitem{Saxt}
C.~J. Saxton,
Galaxy stability within a self-interacting dark matter halo,
Mon. Not. R. Astron. Soc. \textbf{430}, 1578 (2013).

\bibitem{SaxWu}
C.~J. Saxton, K. Wu,
Gravitational and distributed heating effects of a cD galaxy on the hydrodynamical structure of its host cluster,
Mon. Not. R. Astron. Soc. \textbf{437}, 3750 (2014).

\bibitem{SaxSorWu}
C.~J. Saxton, R. Soria, and K. Wu,
Dark halo microphysics and massive black hole scaling relations in galaxies,
Mon. Not. R. Astron. Soc. \textbf{445}, 3415 (2014).

\bibitem{CRMNSZ}
L.~G. Cabral-Rosetti, T. Matos, D. N\'{u}\~{n}ez, R. A. Sussman, and J. Zavala,
Stellar polytropes and Navarro–Frenk–White dark matter halos: a connection to Tsallis entropy,
arXiv:astro-ph/0405242.

\bibitem{CFSBSRG}
J. Calvo, E. Florido, O. S\'{a}nchez, E. Battaner, J. Soler, and B. Ruiz-Granados,
On a unified theory of cold dark matter halos based on collisionless Boltzmann–Poisson polytropes,
Physica A \textbf{388}, 2321 (2009).

\bibitem{FerHjo}
C. F\'{e}ron, J. Hjorth,
Simulated dark-matter halos as a test of nonextensive statistical mechanics,
Phys. Rev. E \textbf{77}, 022106 (2008).

\bibitem{Peebl}
P.~J.~E. Peebles,
Fluid dark matter,
Astrophys. J. \textbf{534}, L127 (2000).

\bibitem{Goodm}
J. Goodman,
Repulsive dark matter,
New Astron. \textbf{5}, 103 (2000).

\bibitem{HarMoc}
T. Harko, G. Mocanu,
Cosmological evolution of finite temperature Bose-Einstein condensate dark matter,
Phys. Rev. D \textbf{85}, 084012 (2012).

\bibitem{DwKerGer}
M. Dwornik, Z. Keresztes, and L.~A. Gergely,
Rotation curves in Bose-Einstein condensate dark matter halos,
arXiv:1312.3715.

\bibitem{LeonWood}
K.~E. Leonard, R.~P. Woodard,
Graviton corrections to vacuum polarization during inflation,
Class. Quantum Grav. \textbf{31}, 015010 (2014).

\bibitem{GMPW}
D. Glavan, S.~P. Miao, T. Prokopec, and R.~P. Woodard,
Electrodynamic effects of inflationary gravitons,
Class. Quantum Grav. \textbf{31}, 175002 (2014).

\bibitem{WanWood}
C.~L. Wang, R.~P. Woodard,
Excitation of photons by inflationary gravitons,
arXiv:1408.1448.

\bibitem{MarPrad}
A.~L. Maroto, F. Prada,
Higgs effective potential in a perturbed Robertson-Walker background,
Phys. Rev. D \textbf{90}, 123541 (2014).

\bibitem{Barrow}
J.~D. Barrow,
Graduated inflationary universes,
Phys. Lett. B \textbf{235}, 40 (1990).

\bibitem{Chavan}
P.-H. Chavanis,
Models of universe with a polytropic equation of state I. The early universe,
Eur. Phys. J. Plus \textbf{129}, 38 (2014).

\bibitem{KleidSpy}
K. Kleidis, N.~K. Spyrou,
Polytropic dark matter flows illuminate dark energy and accelerated expansion,
arXiv:1411.6789.

\bibitem{ChernJETP}
A.~D. Chernin,
Dark energy in systems of galaxies,
Pis'ma Zh. Eksp. Teor. Fiz. \textbf{98}, 394 (2013)
[JETP Lett. \textbf{98}, 353 (2013)].

\bibitem{LukRub}
V.~N. Lukash, V.~A. Rubakov,
Dark energy: myths and reality,
Phys. Usp. \textbf{51}, 283 (2008).

\bibitem{Ven}
A.~E.~M. van de Ven,
Index-free heat kernel coefficients,
Class. Quantum Grav. \textbf{15}, 2311 (1998).

\bibitem{Fursaev1}
D.~V. Fursaev,
Kaluza-Klein method in theory of rotating quantum fields,
Nucl. Phys. B \textbf{596}, 365 (2001).

\bibitem{Fursaev2}
D.~V. Fursaev,
Statistical  mechanics,  gravity,  and  Euclidean  theory,
Nucl. Phys. B (Proc. Suppl.) \textbf{104}, 33 (2002).

\bibitem{Bhatt1}
S. Bhattacharyya,
Constraints on the second order transport coefficients of an uncharged fluid,
JHEP \textbf{1207}, 104 (2012).

\bibitem{Bhatt2}
S. Bhattacharyya,
Entropy current from partition function: one example,
JHEP \textbf{1407}, 139 (2014).

\bibitem{Bhatt3}
S. Bhattacharyya,
Entropy current and equilibrium partition function in fluid dynamics,
JHEP \textbf{1408}, 165 (2014).

\bibitem{ZhuKlau}
C. Zhu, J.~R. Klauder,
The classical limit of ultralocal scalar fields,
J. Math. Phys. \textbf{35}, 3400 (1994).

\bibitem{LandLifshElas}
L.~D. Landau, E.~M. Lifshitz,
\textsl{Theory of Elasticity}
(Pergamon, New York, 1970).

\bibitem{Migdal}
A.~B. Migdal,
Vacuum polarization in strong fields and pion condensation,
Sov. Phys. Usp. \textbf{20}, 879 (1977).

\bibitem{Newton_scat}
R.~G. Newton,
\textsl{Scattering Theory of Waves and Particles}
(Springer-Verlag, New York, 1982).

\bibitem{Eckart}
C. Eckart,
Vortices and streams caused by sound waves,
Phys. Rev. \textbf{73}, 68 (1948).

\bibitem{Wester}
P.~J. Westervelt,
The theory of steady rotational flow generated by a sound field,
J. Acoust. Soc. Am. \textbf{25}, 60 (1953).

\bibitem{RudSol}
O.~V. Rudenko, S.~I. Soluyan,
\textsl{Theoretical foundations of nonlinear acoustics}
(Consultants Bureau, London, 1977)

\bibitem{Lighthill}
J. Lighthill,
Acoustic streaming,
J. Sound Vib. \textbf{61}, 391 (1978).

\bibitem{CorJaTom}
J.~M. Cornwall, R. Jackiw, and E. Tomboulis,
Effective action for composite operators,
Phys. Rev. D \textbf{10}, 2428 (1974).

\bibitem{BjoDre}
J.~D. Bjorken, S.~D. Drell,
\textsl{Relativistic Quantum Theory} Vol. I: \textsl{Relativistic Quantum Mechanics}
(McGraw-Hill, New York, 1964).

\bibitem{WeinbergB}
S. Weinberg,
\textsl{The Quantum Theory of Fields} Vol. 1: \textsl{Foundations}
(Cambridge University Press, Cambridge, 1996).

\end{thebibliography}
\end{document}